\documentclass[pra,letterpaper,onecolumn,showpacs,superscriptaddress]{revtex4}
\usepackage{graphicx,psfrag,amsmath,amssymb,amsfonts,bbm,latexsym,color,dcolumn}

\usepackage{graphicx,psfrag,amsmath,amssymb,amsfonts,bbm,latexsym,color,dcolumn,bm}

\begin{document}
\title{On electrostatic and Casimir force measurements
between conducting surfaces in a sphere-plane configuration}

\author{W.J. Kim$^*$}

\affiliation{Department of Physics and Astronomy, Dartmouth College,
6127 Wilder Laboratory, Hanover, NH 03755, USA}

\author{M. Brown-Hayes}

\affiliation{Department of Physics and Astronomy, Dartmouth College,
6127 Wilder Laboratory, Hanover, NH 03755, USA}

\author{D.A.R. Dalvit}

\affiliation{Theoretical Division, MS B213, Los Alamos National Laboratory,
Los Alamos, NM 87545, USA}

\author{J.H. Brownell}

\affiliation{Department of Physics and Astronomy, Dartmouth College,
6127 Wilder Laboratory, Hanover, NH 03755, USA}

\author{R. Onofrio}

\affiliation{Dipartimento di Fisica ``Galileo Galilei'', Universit\`a  di Padova,
Via Marzolo 8, Padova 35131, Italy}

\affiliation{Department of Physics and Astronomy, Dartmouth College,
6127 Wilder Laboratory, Hanover, NH 03755, USA}

\date{\today}

\begin{abstract}
We report on measurements of forces acting between two conducting
surfaces in a spherical-plane configuration in the 35 nm-1 $\mu$m
separation range. The measurements are obtained by performing
electrostatic calibrations followed by a residuals analysis after
subtracting the electrostatic-dependent component. 
We find in all runs optimal fitting of the calibrations for exponents 
smaller than the one predicted by electrostatics for an ideal
sphere-plane geometry. We also find that the external bias potential 
necessary to minimize the electrostatic contribution depends on the 
sphere-plane distance. In spite of these anomalies, with a proper 
but model-dependent subtraction of the electrostatic contribution 
we have found evidence for short-distance attractive forces of magnitude 
comparable to the expected Casimir-Lifshitz force. 
We finally discuss the relevance of our findings in the more 
general context of Casimir-Lifshitz force measurements, with 
particular regard to the critical issues of the electrical 
and geometrical characterization of the involved surfaces. 
\end{abstract}


\maketitle

\section{Introduction}
Casimir-Lifshitz forces \cite{Casimir,Lifshitz} provide an
experimental window on the nature of quantum vacuum at the macroscopic
level \cite{Plunien,Milonni,Mostepanenko,Bordag,Bordag1,Reynaud,Milton,Milton1,Lamoreauxrev}.
After pioneering attempts in the original parallel plate configuration
\cite{Sparnaay}, and in a variant of this configuration based upon a
sphere and a plane \cite{vanBlockland}, the claimed accuracy of recent
experiments ranges from  15$\%$ in the parallel plane case
\cite{Bressi} to 0.2-5$\%$ in the sphere-plane case
\cite{Lamoreaux,Mohideen,Chan,Decca,SuperDecca}. In the case of the sphere-plane
configuration, the Casimir force is typically evaluated by using the so-called
proximity force approximation (PFA) \cite{Derjaguin, Blocki},
introducing a theoretical uncertainty  estimated to be in the 0.1 $\%$ 
range, and recently investigated experimentally \cite{DeccaPFA}.

A sphere-plane experiment has been performed, with a radius of
curvature for the sphere of order of cm, at relatively large
distances above 1$\mu$m \cite{Lamoreaux}. In this case
the largest deviation from the bare Casimir formula for an idealized
configuration (perfect conductors, zero temperature) is expected to be
the thermal contribution to the radiation pressure  on the surfaces,
still below the sensitivity of the apparatus at  the largest
explored distances. Three sphere-plane experiments have also been
performed in a quite distinct regime, with smaller radius of
curvature, of order 100 $\mu$m, and in the range below one micrometer
\cite{Mohideen,Chan,Decca,Decca1}. 
At distances less than 1 $\mu$m the correction to the Casimir force 
due to finite conductivity and surface roughness of the substrates 
cannot be neglected at the level of accuracy claimed in these three 
experiments. At least other five groups have recently performed Casimir 
force experiments in the sphere-plane configuration 
\cite{Palasantzas,Chevrierpaper,Petrov,Ludwig,Chanprl}.

Mastering the Casimir force at the highest level of accuracy is
essential to provide reliable limits to other macroscopic forces
acting in the micrometer range, such as corrections to the Newtonian
gravitational force independently predicted in many attempts to unify
gravitation with the remaining fundamental interactions
\cite{Fishbach,Gundlach,Onofrio}. Although in the nanometer range the
Casimir force loses its universal nature morphing into more specific,
structure-dependent, molecular van der Waals forces, its study is
sometimes considered of some relevance for designing nanomechanical
structures and to investigate nonlinear effects, as first discussed in
\cite{Serry} and then experimentally demonstrated in \cite{Buks,Chan1}.

In this paper we report more extensively on measurements performed in a sphere-plane
geometry exploring a novel range of parameters that interpolates
between the two previous sets of sphere-plane measurements
\cite{KimPRARC}. We use a gold-coated spherical lens with large
radius of curvature, similar to the experiment performed by Steve
Lamoreaux in Seattle, while at the same time exploring distances down
to few tens of nanometers from the point of contact between the sphere
and the plane, similar to more microscopic setups using
microresonators. While we do achieve some evidence for the expected
Casimir-Lifshitz force, our measurements suggest that a reanalysis of
systematic effects in previous experiments will be beneficial to
assess the accuracy with which this force has been measured so far,
with particular regard to the role of the residual electrostatic
forces \cite{Stipe,Speake1}.

The paper is organized as follows: in Section II we describe the
experimental apparatus and the basics of our measurement technique. In
Section III we describe a procedure for electrostatic calibrations
to determine various critical parameters such as the calibration
factor, the absolute separation, and the minimizing potential related to
the contact potential. Careful control over the system parameters
allows us to search for a non-Coulombian contribution to the observed 
force signal and therefore to test this residual against various 
hypothesis such as a possible uncompensated electrical voltage, and 
the presence of the Casimir-Lifsthiz force, as described in Section
IV. In Section V we critically assess systematic effects in our 
measurements and compare our findings to both previous short-range 
Casimir force measurements and long-range atomic force microscopy measurements.

\vspace{0.5cm}
\begin{figure}[t]
\begin{center}
\includegraphics[width=0.8\columnwidth]{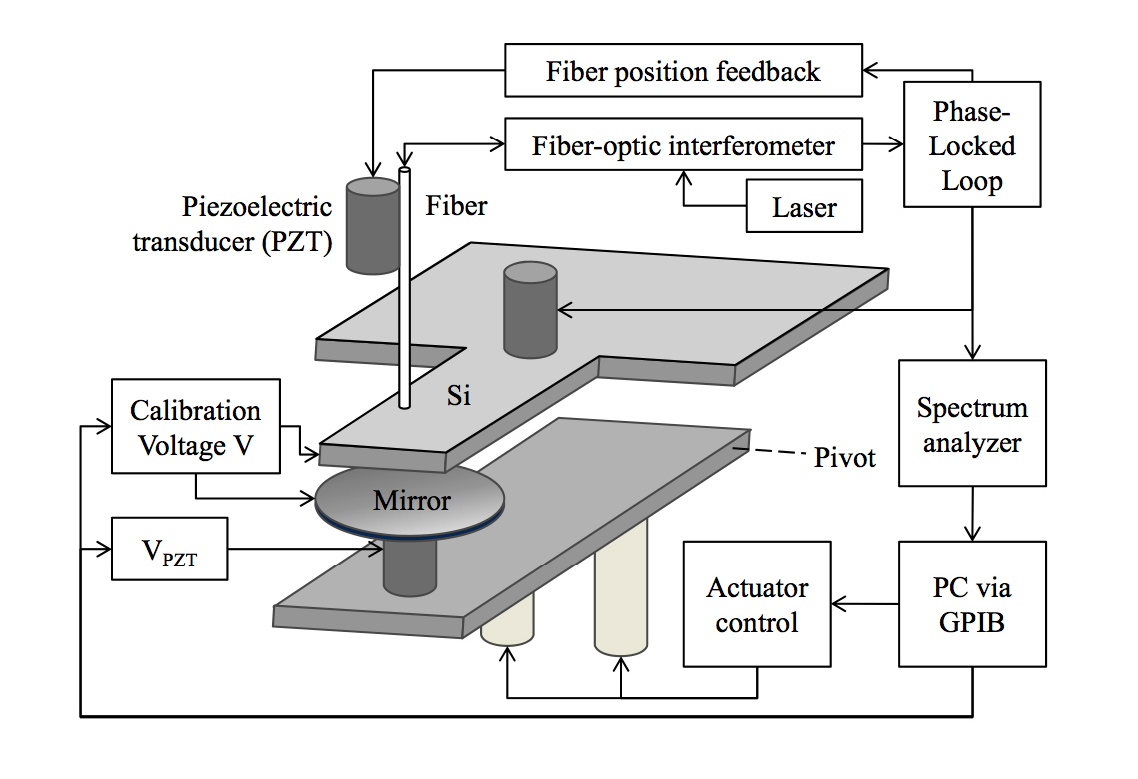}
\end{center}
\caption{Schematic of the experimental setup and the data acquisition
  scheme. The silicon resonator is opposed by a spherical mirror, both
  Au coated. The latter can be moved along the vertical direction by
  driving two mechanical actuators for coarse approach, and with a
  piezoelectric actuator for fine control of the separation. The
  outside face of the resonator reflects light coming from a fiber
  optic interferometer, located 20-100 $\mu$m above it. The resonator
  is driven weakly at its resonant frequency with a piezoelectric
  actuator connected in a phase-locked loop (PLL) to the optical fiber.}
\end{figure}

\begin{figure}[t]
\vspace{1.0cm}
\begin{center}
\includegraphics[width=0.50\columnwidth]{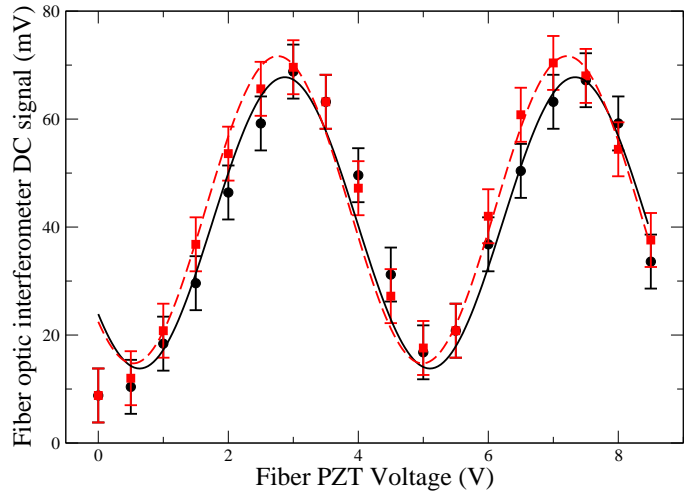}
\end{center}
\caption{Calibration of the PZT actuator governing the sphere-plane
  separation. Four calibrations of the PZT actuator have been performed during the
  overall data taking, of which only two are reported here for
  simplicity, indicated with red-square and black-dot points, and their best fit with a 
  sinusoidal function (red, dotted line, and black, continuous line,
  respectively). The average value of the actuation coefficient from the
  four data sets is $\beta=(87 \pm 2)$ nm/V.}
\end{figure}

\section{Experimental set-up}

The experimental setup is an upgrade of the one already described in a
previous paper  \cite{Brown} for measuring the Casimir force in the
cylinder-plane configuration as first theoretically discussed in \cite{eccentric}. 
Forces acting between a flat surface and a spherical
surface are detected by measuring the shift in the mechanical
oscillation frequency of a silicon cantilever that acts as the plane
surface. The main components in this setup are the use of a
phase-locked loop scheme \cite{Rugar} to drive the cantilever at its
resonant frequency, the use of optical quality surfaces for both the
resonator and the sphere, and a faster and more flexible data
acquisition scheme.

A schematic of the apparatus is shown in Fig. 1. 
The rectangular cantilever, of length $L=(22.56 \pm 0.01)$ mm, width
$w=(9.93 \pm 0.01)$ mm and thickness $t=(330 \pm 10)$ $\mu$m was
laser-cut from a silicon wafer. Given the density of silicon,
$\rho=2.3 \times 10^3$ kg/m$^3$, the physical mass of the cantilever
is $m_p=(1.72 \pm 0.05) \times 10^{-4}$ kg. The cantilever is
sandwiched between two aluminum clamping structures, electrically
isolated and thermally stabilized by a Peltier cooler. This last
feature is required since we have observed that the resonant frequency
of the cantilever drifts as much as 2 Hz as the ambient temperature
fluctuates about 1 K over the duration of a typical run of a few
hours. The effect of the thermal drifts has also been mitigated
off-line by a proper implementation of the data acquisition sequence,
as described in Section III.

Below the cantilever a spherical mirror with radius of curvature
$R=(30.9 \pm 0.15)$ mm and diameter $a=(8.00 \pm 0.25)$ mm is mounted
on an aluminum frame connected to two motorized actuators allowing for
coarse translational motion, plus an additional piezoelectric
transducer for fine translational motion driven with a bias
$V_{\mathrm{PZT}}$. The spherical lens is gold-coated by the 
manifacturing company with a thickness of 250 nm ($\pm$ 10 $\%$)
The surface of the cantilever facing the spherical lens is coated 
with a 200 nm layer of gold ($\pm$ 10 $\%$) by thermal evaporation, 
with the rate of evaporation kept below 50 $\AA$/sec at a pressure 
of $3 \times 10^{-6}$ mbar to ensure better conditions for homogeneous coating.

The predicted frequency of the fundamental flexural mode of the
cantilever is $\nu_p= (0.162 t/L^2) \sqrt{E/\rho}$ with $E$ the
Young's modulus of the material \cite{Timoshenko}. In our case this
yields a resonant frequency around 894 Hz, within few percent from the
measured frequency at the typical vacuum pressure of $1.6 \times
10^{-4}$ Torr in the vacuum chamber. 
The stiffness $k$ of the resonator can be estimated by
assuming an effective mass for the resonator mode roughly equal to the
physical mass, which yields $k=1.036 E w t^3/L^3 \simeq 5.4 \times
10^3$ N/m. The stiffness is three or four orders of magnitude higher
than the typical cantilevers used in atomic force microscopy, implying
both advantages and drawbacks. A large stiffness allows us to reliably
achieve small gaps with little static deflection of the cantilever in
the presence of strong electric fields  for calibration purposes, 
at the expense of a lower force sensitivity.  By decreasing the
stiffness of the cantilever both the reachable minimum gap before snapping 
between the two surfaces occurs and the largest explorable gap for 
a given signal to noise ratio are increased.

The motion of the cantilever is detected by using a fiber optic
interferometer \cite{Rugar1} positioned a few microns above the upper
cantilever face and fed by temperature-stabilized diode laser with an
adjustable power in the 5-10 mW range at the wavelength of 781 nm. The
output signal from the interferometer is filtered and amplified
through a single reference mode lock-in amplifier and is fed back into
the piezoelectric actuator driving the cantilever motion. The phase
between the input and the output signals is properly chosen to
maximize the vibration amplitude, typically around 30-40 degrees. This
scheme is much faster and more efficient than the previous open-loop
scheme as described in \cite{Brown} in which white noise was used as
the driving source without a feedback, and the complete FFT spectrum
was acquired. The vibration frequency was measured by a counter with
10 mHz  resolution, 100 times faster than in the open-loop scheme.
The front side of the chamber has a large viewport allowing for visual 
inspection of the relevant components through an optical microscope,
as shown in Fig. 10 of \cite{Brown}. This also provides for a quick 
assessment of the optical fiber location, which can be manually
adjusted within few hundred micrometers from the underlying resonator 
using a feedthrough micrometer. 

The PZT was calibrated by using the same fiber optic
interferometer with the cantilever removed and the spherical mirror
replaced by a flat mirror. Assuming a linear relationship $\Delta x =
\beta \Delta V_{\rm PZT}$, the DC interference amplitude versus the
voltage applied to the PZT, as shown in Fig. 2, was fit with a
sinusoidal function and the distance inferred from its period to be half of 
the 781 nm laser wavelength.  The average of four separate calibration 
runs yields $\beta=(87 \pm 2)$ nm/V, and it was found to be independent 
of the voltage applied to the PZT, within the error, in the 0-100 V range. 
The sensitivity of the fiber optic interferometer is evaluated in the 
same configuration by driving the PZT with a monoschomatic signal at
various amplitudes, and looking the corresponding signal at the FFT 
spectrum analyzer. When the driving amplitude at a frequency of 1 KHz 
(still well below the maximum response frequency of the PZT of 69 KHz) is
reduced at the level of a peak-to-peak signal of 0.25 mV, the 
measured amplitude at the FFT spectral analyzer becomes equal to the 
FFT broadband noise level of 10 $\mu V_{\mathrm{rms}}$ with an integration time 
of 10 s. Using the determined value of the actuation coefficient 
of the PZT, we estimate the minimum detectable signal with SNR $\simeq
1$ to be $\simeq 0.80 \AA/\sqrt{\mathrm{Hz}}$, similar to the one
reported in \cite{Rugar1}.

For a generic distance-dependent force and for a small amplitude of
the cantilever oscillation (2-3 nm as inferred from the interferometer
signal) with respect to the average separation between the two
surfaces $x$, the measured frequency of the resonator $\nu_m$ is shifted 
with respect to the proper frequency of the resonator $\nu_p$ in such
a way that
\begin{equation}
\Delta \nu^2= \nu_m^2 -\nu_p^2 = -\frac{1}{4 \pi^2 m_\mathrm{eff}}
\frac{\partial F(x)}{\partial  x},
\end{equation}
where $m_\mathrm{eff}$ is the effective mass of the mode of
oscillation of the resonator.  In general, the total force acting on
the resonator is the aggregate from independent sources and the square of
the measured frequency $\nu_m$ of a cantilever can be identified as
the sum of contributions from those sources,
\begin{equation}
\nu_m^2(x,V) = \nu_p^2 + \Delta\nu_e^2(x,V) + \Delta\nu^2_r(x),
\end{equation}
where $\Delta\nu_e^2$ is the frequency shift due to electrical force
gradients for instance due to an external bias voltage $V$, and 
$\Delta\nu^2_r$ is the frequency shift subject to force gradients of
non-electrostatic nature, for instance the Casimir force, or random drifts 
of instrumental and environmental origin. 
In the next section we will discuss the response of the resonator 
in the case of electrostatic forces.

\section{Electrostatic calibrations}

We start with the electrostatic energy stored in a generic capacitance 
$C(x)$ biased by an external potential $V$, and under the presence of 
a contact potential $V_c(x)$ which in general may depend on the distance: 

\begin{equation}
E_{el}=\frac{C(x)}{2}[V-V_c(x)]^2.
\end{equation}

As recently pointed out by Lamoreaux \cite{Lamoreaux2}, if the contact
potential depends on distance the electrostatic force will include 
an additional term independent upon the external bias voltage.
Since in our configuration the observable is the frequency shift, 
proportional to the second derivative of the electrostatic energy with 
respect to $x$, we have, after regrouping the various terms in powers of $V-V_c$:
\begin{equation} 
E_{el}'' = \frac{C''}{2}(V-V_c)^2-[2 C' V'_c+C V''_c](V-V_c)+C {V'_c}^2.
\end{equation}
This can be regrouped again as $E_{el}'' = A(V-V_c+B)^2+D$ and, identifying 
the coefficients $A, B, D$ to match the terms in Eq. 4, we obtain
\begin{equation}
A=\frac{C''}{2} ~ ~, ~ ~ ~
B=-\frac{2 C' V'_c+C V''_c}{C''} ~ ~, ~ ~ ~ 
D=C V_c'^2 - \frac{(2 C' V'_c+C V''_c)^2}{2C''}.
\end{equation}
This allows to write:
\begin{equation}
E_{el}'' = \frac{C''}{2}(V-V_0)^2+C{V'_c}^2-\frac{(2 C' V'_c+C V''_c)^2}{2C''},
\end{equation}
\noindent
where
\begin{equation}
V_0(x)=V_c(x)+\frac{2 C' V'_c+C V''_c}{C''}
\label{V_0diffequ}
\end{equation}
\noindent
is the {\it minimizing} external potential, {\it i.e.} the external
bias minimizing the electrostatic contribution to the frequency shift.
It is clear that, even if the external bias voltage is chosen to
minimize the electrostatic contribution, an irreducible term 
containing the first and second spatial derivatives of $V_c$ 
in Eq. 7 still causes a frequency shift. 
The expected frequency shift due to Coulombian interactions is therefore:
\begin{equation}
\nu^2 = {\nu_0}^2- \frac{C''}{8 \pi^2 m_\mathrm{eff}}(V-V_0)^2+
\frac{1}{4 \pi^2 m_\mathrm{eff}}
\left[-C{V'_c}^2+\frac{(2 C' V'_c+C V''_c)^2}{2C''}\right].
\end{equation}
We identify three contributions on the RHS of Eq. 8: 
the first term is unrelated to external or internal voltages, the second
term depends on the external voltage, and the last term is 
a Coulombian contribution only related to the spatial dependence 
of the contact potential. The electrostatic calibrations
consist in the study of the response of the cantilever frequency to 
an external potential, therefore selecting the second contribution
alone which allows to infer both the effective mass of the resonator 
and the minimizing potential $V_0$. This knowledge is then used  
to infer the contact potential assuming a specific form of the 
functional dependence on distance of $V_0$, which is in turn 
used to identify the third contribution in Eq. 8. 
The subtraction of the total Coulombian contribution allows then to 
evaluate the square frequency related to non-Coulombian forces and to  
random drifts due to background noise in the apparatus (see section
IV) as first discussed in AFM measurements by \cite{Garcia} and in 
the specific context of Casimir measurements in \cite{Iannuzzi}.

\begin{figure}[t]
\vspace{1.0cm}
\begin{center}
\includegraphics[width=0.70\columnwidth]{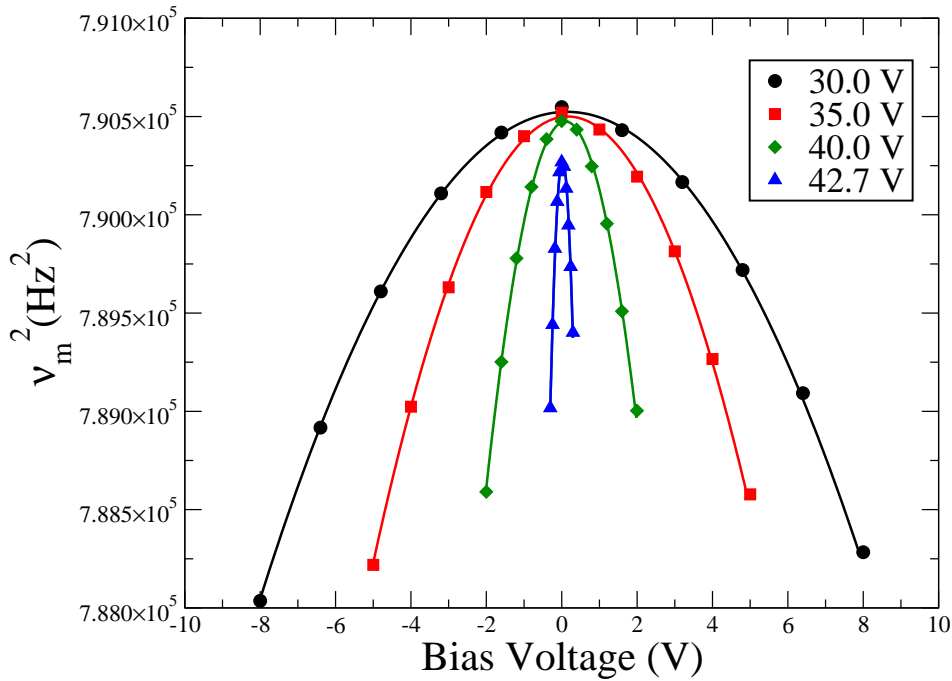}
\end{center}
\caption{Parabolic dependence of the squared frequency $\nu_m^2$ upon the
external bias voltage $V$ for a typical experimental run.
The different curves correspond to
different values of the sphere-plane separation (defined in the legend
in terms of the voltage applied to the PZT used for approaching the sphere).
The range of external bias voltages is adapted for each sphere-plane 
separation to maintain comparable values of the maximum detected
frequency shift, of about 1 Hz. This corresponds to keep a constant 
value for the electric field in the gap.  
The best fit with a parabolic function is also shown, with the
parabolic curves relative to the largest PZT voltages (corresponding
to smaller gap separations) displaying largest curvatures. 
No significant dependence of the fitting parameters has been observed 
by changing the external bias voltage span.}
\end{figure}

\begin{figure*}[t]
\vspace{1.0cm}
\begin{center}
\includegraphics[width=0.80\textwidth]{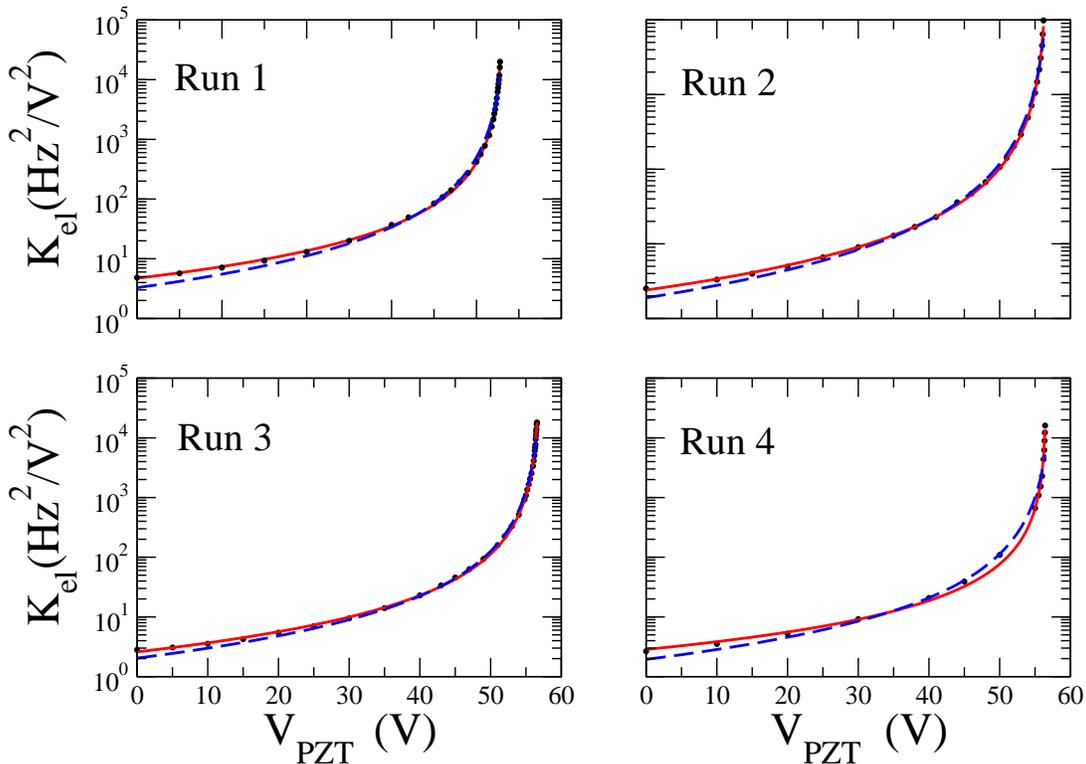}
\end{center}
\caption{Electrostatic curvature coefficient versus the voltage
applied to the PZT actuator for four different runs, and related fits 
with the expected fixed exponent $e=-2$ (blue dashed line) as in Eq. 10
and by allowing the exponent to be a free parameter (red continuous line).
In run 1  the location of the surface of the sphere in proximity
of the cantilever is different from the other runs, as the latter
were taken after tilting the sphere by an angle equal to
5$\times$10$^{-2}$ radians.}
\end{figure*}

Regarding the capacitance, for the sphere-plane configuration and for 
our choice of parameters, the proximity force approximation (PFA)
holds and the capacitance is $C(x) =  2 \pi \epsilon_0 R \ln(R/x)$.  
This formula in principle must be corrected for the finite spherical 
mirror diameter used in alternative to a full sphere, but this is a 
sub-leading PFA correction approximately equal to $0.1 \%$
for a typical separation of $d=1 \mu{\rm m}$. 
Therefore, the measured frequency $\nu_m$ of the cantilever 
can be parameterized as
\begin{equation}
\nu_m^2(x) = \nu_0^2(x) - K_{\mathrm{el}}(x)(V - V_0)^2 ,
\end{equation}
where $\nu_0^2(x)=\nu_p^2+\Delta\nu_e^2(x,V_0)+\Delta \nu_r^2(x)$, a parabola
whose maximum is reached when the applied voltage at a given distance
equals to the minimizing potential $V_0$. The parabola curvature 
$K_{\mathrm{el}}(x) = \epsilon_0 R / 4 \pi m_{\rm eff} x^2$ reflects 
the cantilever response to externally applied electric forces at 
a given distance $x$ and allows to extract the effective mass of the
resonator mode, and it can be measured by sweeping the applied bias 
over a bipolar range to discern the parabolic dependence of the 
frequency shift.  The minimizing potential $V_0$ is found by looking 
at the horizontal displacement of the parabola. 
In Fig. 3 we show examples of electrostatic calibrations in which the square 
resonator frequency is plotted versus the bias voltage for different values 
of the gap separation between the sphere and the plane.
In order to minimize the resonator frequency drifts due to
fluctuations of ambient temperature, we measure the unbiased frequency before
and after each measurement taken at a given bias voltage $V$.
The average of the two unbiased frequencies is then used as the reference to evaluate the shift.
We notice two remarkable features in Fig. 3. First, the peak value of
the frequency is different in the four cases, reflecting the possible 
presence of a distance-dependent force combined with possible drifts 
in the value of the intrinsic frequency. 
Second, the location of the peak value of the resonator frequency 
in the parabolic fit does not occur at the same value of external voltage,
indicating a distance-dependent minimizing potential.

\begin{figure}[t]
\vspace{3.0cm}
\begin{center}
\includegraphics[width=0.80\columnwidth]{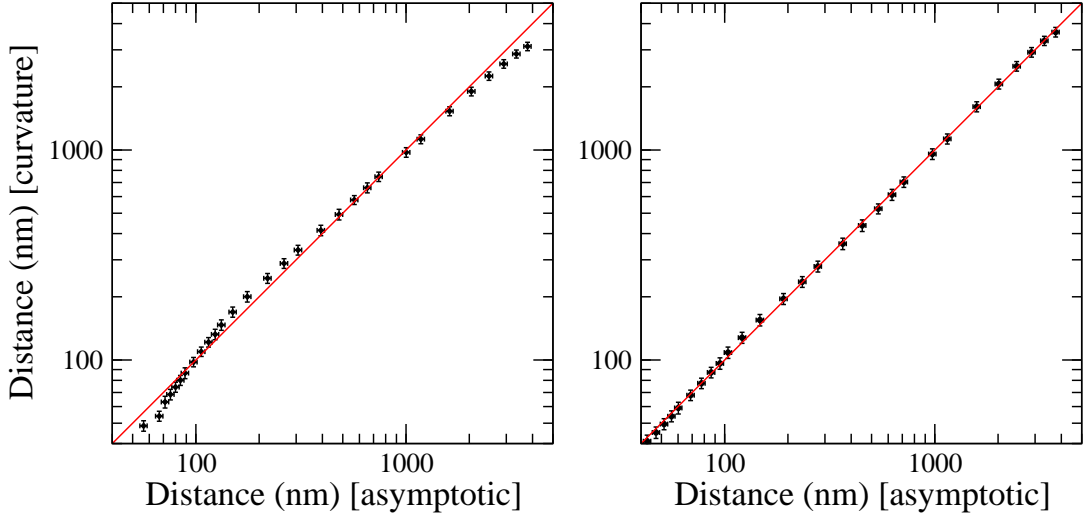}
\end{center}
\caption{Plot of $x(V_{\rm PZT})$ converted into distance, as
evaluated from the calibration factor $\alpha$ for the inverse 
square law, versus the same function evaluated from the asymptotic 
limit of the fit function. The left plot uses the expected 
electrostatic law with an exponent $e=-2$, while the right plot 
is obtained by optimizing the exponent in the curvature scaling 
with distance. Red lines are added to provide an eyeguide for 
the deviations from the expected consistency between the 
two determinations of the same quantity.}
\end{figure}

The absolute distance between the sphere and the plane is not known
with sufficient accuracy prior to the force measurement. In the
experiment, the gap is varied by the voltage applied to the PZT
($V_{\rm PZT}$) and consequently $K_{\rm el}$ is a function of the
relative distance ({\it i.e.}, of the applied $V_{\rm PZT}$). This
requires an additional fitting parameter $V_{\rm PZT}^0$, which would
cause contact and must be inferred from fitting the function
\begin{equation}
\label{kelectrostatic}
K_{\rm el}(V_{\rm PZT})=\alpha (V_{\rm PZT}^0 - V_{\rm PZT})^{-2},
\end{equation}
where $\alpha$ is a calibration factor containing the effective mass 
of the cantilever through $\alpha \equiv \epsilon_0 R / 4 \pi m_{\rm eff} \beta^2$.
In Fig. 4 the curvature parameter $K_{\mathrm{el}}$ is plotted
versus the piezoelectric actuator voltage $V_{\rm PZT}$ for four different
experimental runs. Assuming the PZT actuation is linear, the absolute
separation for a given $V_{\rm PZT}$ can be inferred in two ways:
$x(V_{\rm PZT})=\beta (V^0_{\rm PZT}-V_{\rm PZT})$ or $x(V_{\rm
  PZT})=\beta [ \alpha/K_{\rm el}(V_{\rm PZT})]^{1/2}$. Therefore, the
absolute distance can be inferred either from the asymptotic limit
$V^0_{\rm PZT}$ of the fit function or from the calibration factor
$\alpha$ of the same function, indicating an interdependency of the
two physical parameters appearing in Eq.~(\ref{kelectrostatic}). Note
that quite large values of $K_{\mathrm{el}}$, of order $-2 \times 10^4
\mathrm{Hz}^2/\mathrm{V}^2$ in runs 1, 3, and 4, have been achieved. 
These large signals serve to accurately constrain the fit parameters 
owing to the high stiffness of the cantilever, which allows us to
explore very small distances between the sphere and the cantilever without stiction.

Comparison of the two methods for determining $x(V_\mathrm{PZT})$ in
Fig. 5 (left plot) reveals disagreement when applying the inverse square law in
Eq.~\ref{kelectrostatic}.  Direct measurement of the sphere-plane capacitance 
versus gap provides a third assessment of $V^0_{\rm PZT}$. 
In Fig. 6 the capacitance versus PZT voltage is shown with the 
fitting curve corresponding to the capacitance between a sphere 
and a plane in the PFA approximation. 

Another issue complicating the electrostatic calibrations is evident 
in Fig. 7 displaying $V_0$ inferred from Fig. 3 versus $V_{\rm PZT}$. 
While the error bars are relatively large at large gap separations, 
they become smaller at smaller gaps, and two distinct dependencies 
are observed. In runs 1 and 2 the minimizing potential seems to depend 
linearly on the gap separation at small separations, while in the 
remaining runs it appears to tend towards a constant value. 
The significance of this dependence in the context of high-precision 
measurements of the Casimir force is discussed in section V.

\begin{figure}[t]
\vspace{1.0cm}
\begin{center}
\includegraphics[width=0.60\columnwidth]{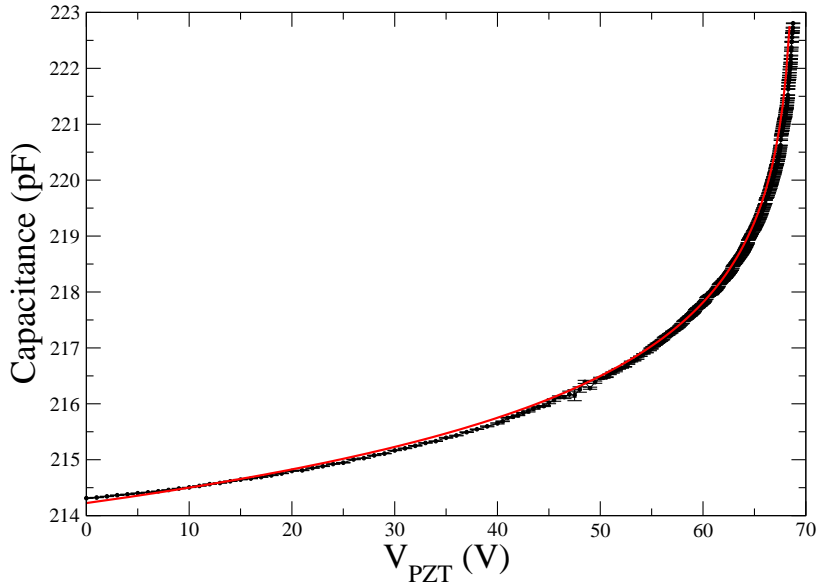}
\end{center}
\caption{Capacitance versus PZT voltage data (black circles) 
and its best fit (red line). The fit is based on the PFA formula 
$C(d)=C_0+ A \ln[\beta(V^0_\mathrm{PZT}-V_\mathrm{PZT})]$, 
where the actuation coefficient $\beta=(87 \pm 2)$ nm/V. 
The best fit yields, for the coefficients $C_0$ and $A$, the 
values $C_0=(193.9 \pm 0.2)$ pF and $A=-(1.757 \pm 0.002)$ pF, 
this last in agreement within two standard deviations with 
the less accurate theoretical expectation  
$A=-2 \pi \epsilon_0 R=-(1.72 \pm 0.02)$ pF. 
The same fit provides a value for the asymptotic value of 
the PZT voltage corresponding to zero gap distance the value 
$V^0_{\rm PZT}=(69.31 \pm 0.01)$ Vm, equivalent to a 
distance of $47 \pm 2$ nm. The best fit corresponds to a 
reduced $\chi^2=2.9$.}
\end{figure}

The error in the curvature parameter $K_{\rm el}$ has been determined
with dedicated runs in which conditions nominally leading to the 
same gap size are maintained, and the result is shown in Fig. 8.
The overall time drift in $K_{\rm el}$, if converted into a thermally induced 
change in gap size, corresponds to $\pm 200$ nm. The average random
uncertainty in $K_{\rm el}$ has been evaluated by taking 
the data of Fig. 8 and subtracting for each point the long term drift by means 
of a moving average with a defined time window. By choosing the average time 
window equal to 4 curvature measurements (each measurement taking 8-10 
minutes depending on the number of points obtained in the frequency versus 
bias voltage measurement) we obtain a relative error in $K_{\rm el}$ equal 
to about 4\% error on top of the fitting uncertainty associated with the parabolic fit. 
This allows us to quantitatively compare fitting functions
for the distance dependence of $K_{\rm el}$, and has resulted in
another anomalous behavior. Indeed, the reduced $\chi^2$ is near one 
when the exponent is fit in the -1.7 to -1.8 range, while it is at
least one order of magnitude higher for fixed -2.0 exponent, as shown in 
Table 1. Adherence of our data to a strict power law from the 
farthest to the closest approach in all four cases implies the drift played little 
role during our data runs, except possibly in the case of Run 4. 
However, we find that the data clearly {\it fail} to follow the 
inverse square law dependence of the electrostatic coefficient upon the 
sphere-plane separation. More quantitatively, if the fitting exponent is left 
as a free parameter, our experimental data from four separate sequences
follow a power law with exponents  $-1.70\pm0.01, -1.77\pm0.02, -1.80\pm0.01,
-1.54\pm0.02$, far from the expected value of -2. When substituted
into Eq.~\ref{kelectrostatic}, these exponent values produce better agreement
between the two methods for determining $x(V_{\rm PZT})$, as shown in
the right plot in Fig. 5.

\begin{table}[b]
\centering
\begin{tabular}{cc|cccc}
&  & Run 1 & Run 2 & Run 3 & Run 4\\
\hline
\hline
&fixed exponent&-2&-2&-2&-2\\
& $\alpha$ $\mathrm{(Hz^2V^{-2}m^2)}$ & $6200\pm98$ & $6197\pm97$ & $6701\pm101$ & $6438\pm156$\\
& $x_0$ (nm) & $64.4\pm1.7$ & $90.5\pm2.6$ & $62.6\pm1.7$ & $93.1\pm2.6$\\
&$\chi^2$/DOF& 15.9 & 7.7 &6.9 &36.5\\
\hline
&free exponent & $-1.70\pm0.01$& $-1.77\pm0.02$ & $-1.80\pm0.01$ & $-1.54\pm0.02$\\
& $\alpha$ $\mathrm{(Hz^2V^{-2}m^{e})}$ & $2805\pm92$ & $3021\pm153$ & $3732\pm144$ & $1415\pm69$\\
& $x_0$ (nm) & $29.6\pm0.9$ & $49.6\pm1.7$ & $35.7\pm1.7$ & $20.0\pm1.7$\\
&$\chi^2$/DOF& 1.0 & 0.8 &1.2 &7.0\\
\hline
\hline
\end{tabular}
\label{fitelectro}
\caption{
Fitting parameters of the electrostatic calibrations, and 
values of the reduced $\chi^2$, as the absolute $\chi^2$ normalized by the 
number of degrees of freedom (DOF) in the fitting for a fixed exponent
equal to 2 as expected from electrostatic considerations, and for a 
free exponent e. Note that the calibration factor $\alpha$
characterizes the sensitivity of our apparatus, yielding the 
effective mass in the case of fitting with the expected 
Coulomb exponent equal to 2. Variations of $\alpha$ from run 
to run are attributable to the rearrangement of the resonator 
clamping system, which strongly affects its effective mass. 
If the exponent is different from the expected value of 2, then the 
dimension of $\alpha$ changes and the effective mass cannot be
properly evaluated.}
\end{table}

\begin{figure*}[t]
\vspace{1.5cm}
\begin{center}
\includegraphics[width=0.90\textwidth]{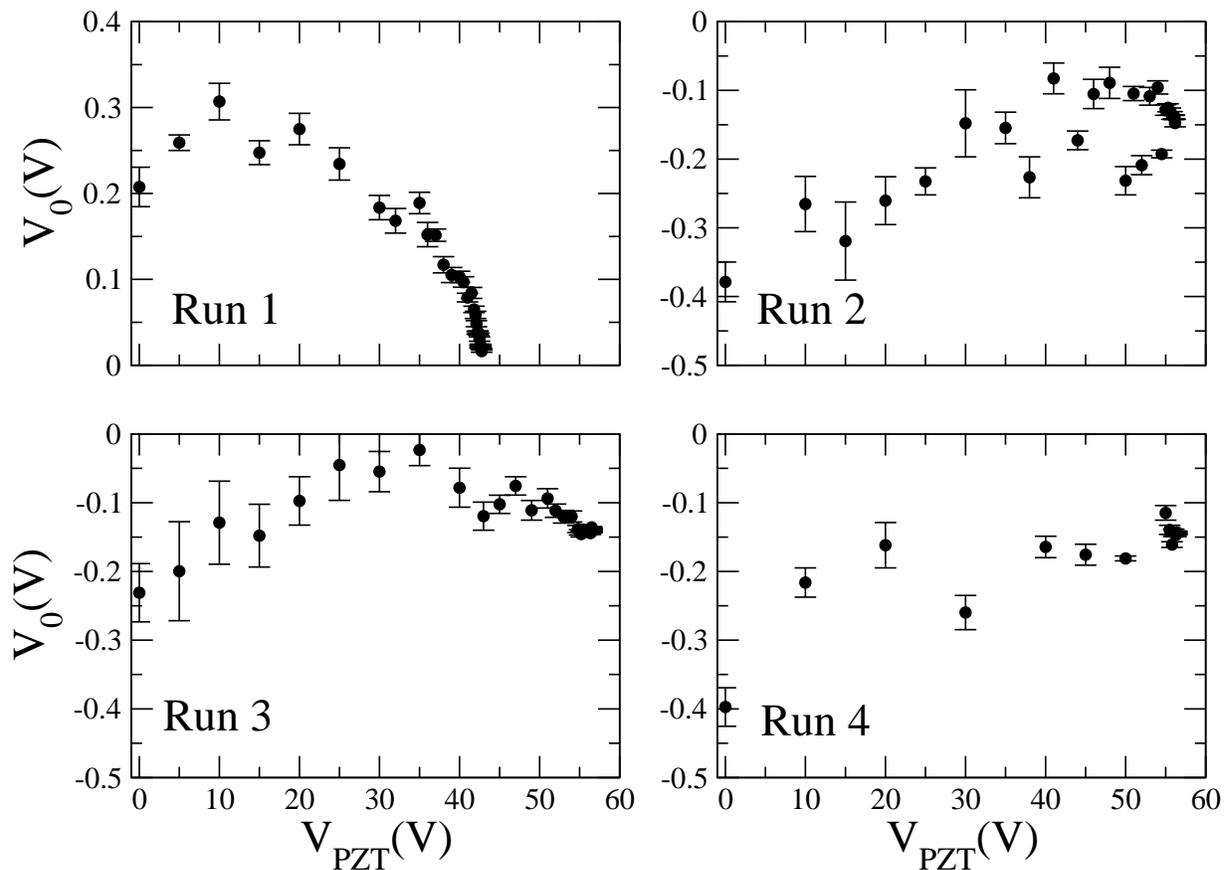}
\end{center}
\caption{Plot of the minimizing potential versus $V_{\rm PZT}$ for  the
  same four runs shown in Fig. 4. Two behaviors are
  evident at small separation (large $V_{\rm PZT}$): either an
  approximately constant potential (Run 3 and Run 4) or a potential
  linearly dependent upon the separation (Run 1 and, to a less extent, 
  Run 2). Note that in run 1 $V_0$ is positive 
  while for Runs 2-4 it is negative and  converges to  $V_0 \simeq$ 
  -150 mV at the smallest separations.}
\end{figure*}

\begin{figure}[t]
\vspace{1.2cm}
\begin{center}
\includegraphics[width=0.80\columnwidth]{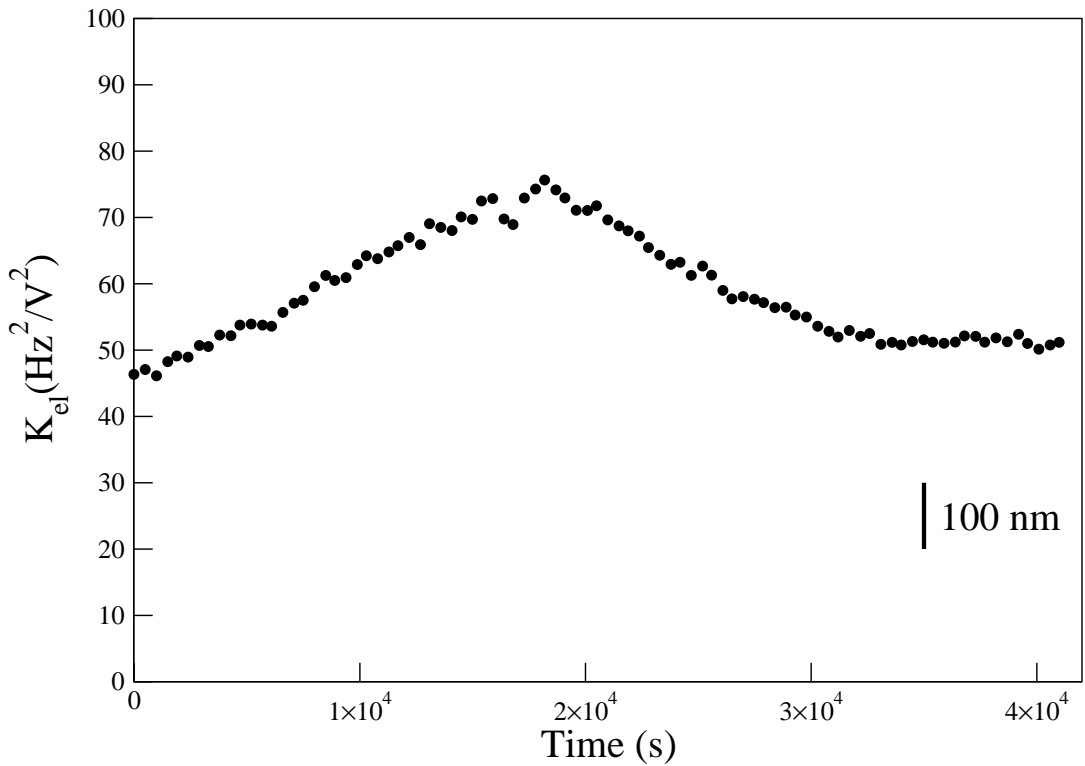}
\end{center}
\caption{Long-term drift of the curvature coefficient $K_{\mathrm{el}}$
during a 12 hours test. The vertical bar represents the equivalent
drift in terms of the change in the sphere-plane separation, in the 
conservative hypothesis that all the drift is attributable to the 
gap drift. Subtracting each point with a window average of 4 measurements 
one gets a relative error in $K_{\rm el}$ of 4 $\%$. The sensitivity 
to the choice of the average window time has been also assessed, with 
relative errors of 2.9 $\%$, 4.1 $\%$, 4.4 $\%$ and 4.6 $\%$ for alternative 
choices of 2, 6, 8, and 10 measurements, respectively.}
\end{figure}

The stability of the electrostatic result has been checked through
repeating the data fit of $K_{\rm el}$ versus $V_{\rm PZT}$,
starting with few points at the largest distances and by progressively
including the data point corresponding to the closer distances.
In this way the intrinsic instability of the fit will
manifest itself through a systematic variation
in the fit parameters subject to a choice of data points included.
Figures 9 and 10 show the result of the stability test for Run 1-4, revealing
dependence of the fits on the number of points considered.
As we will discuss later in Section IV, this is limiting the
accuracy of the determination of the Casimir force.
In Table 1 we report the values of the parameters of
the best fit of the electrostatic calibration for the four runs,
leaving the exponent free, and with the exponent frozen at the
expected value of -2. In the latter case one can infer the effective 
mass of the cantilever and this turns out to be larger by an order 
of magnitude than the physical mass, against the expectation to be 
a fraction of the physical mass \cite{Sader,Cappella,Bressi1}. 
Apart from the uncertainties related to the clamping of the resonator,
this is another signal that the calibration fitting with a pure 
electrostatic contribution neglects some systematic effect.

\begin{figure*}[t]
\vspace{1.5cm}
\begin{center}
\includegraphics[width=0.90\textwidth]{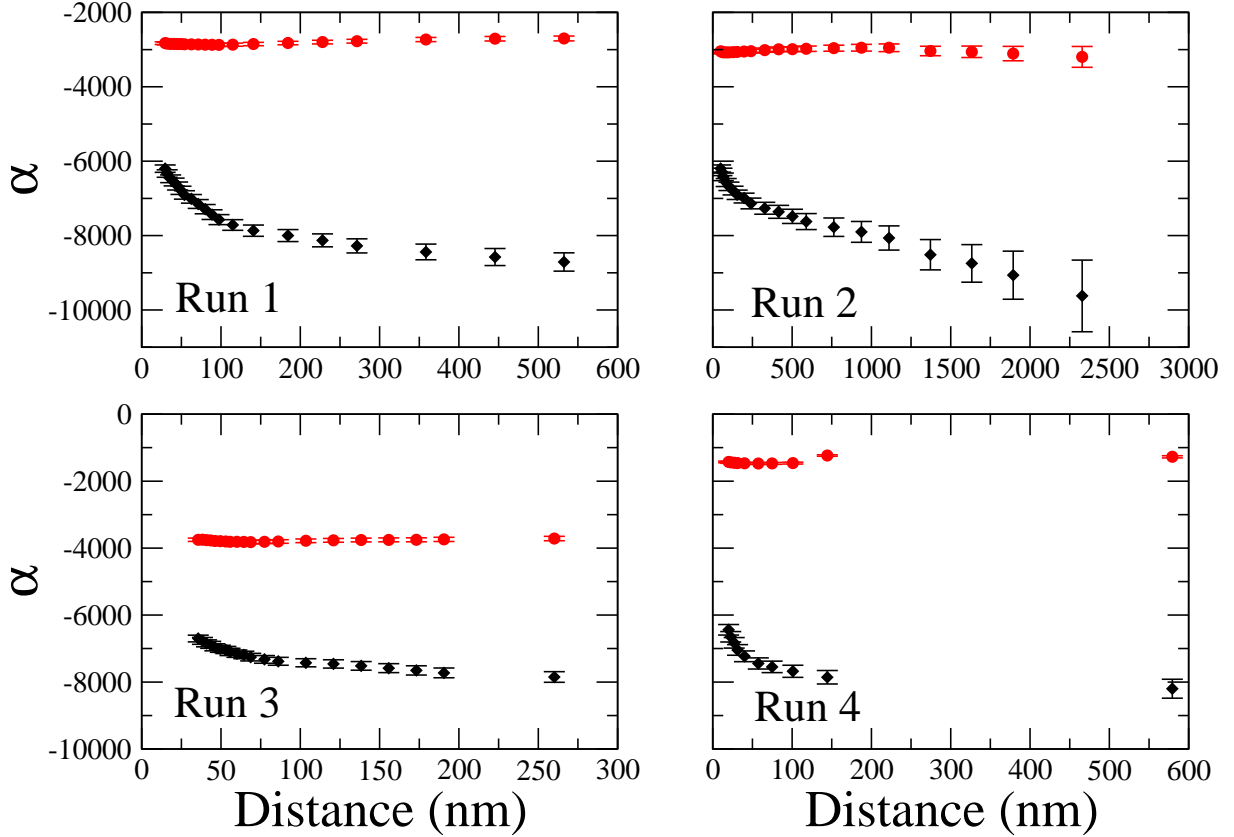}
\end{center}
\caption{Test of stability of the fit of the electrostatic 
calibrations: shown is the determination from the fitting 
of the calibration factor $\alpha$ in the case of the fixed 
-2 exponent versus the sphere-plane distance for the fixed 
-2 exponent (in units of Hz$^2$ V$^{-2}$ m$^2$) expected 
from electrostatics (diamond points, black), and for the best 
exponent left as a free parameter (circle points, red). 
The number of points used in the data fit includes up to 
the gap distance evaluated at that point.  
The best exponent points show convergence to a constant 
value at the closest distances in all four runs, unlike 
the electrostatic case with the fixed exponent.}
\end{figure*}

\begin{figure*}[t]
\vspace{1.5cm}
\begin{center}
\includegraphics[width=0.90\textwidth]{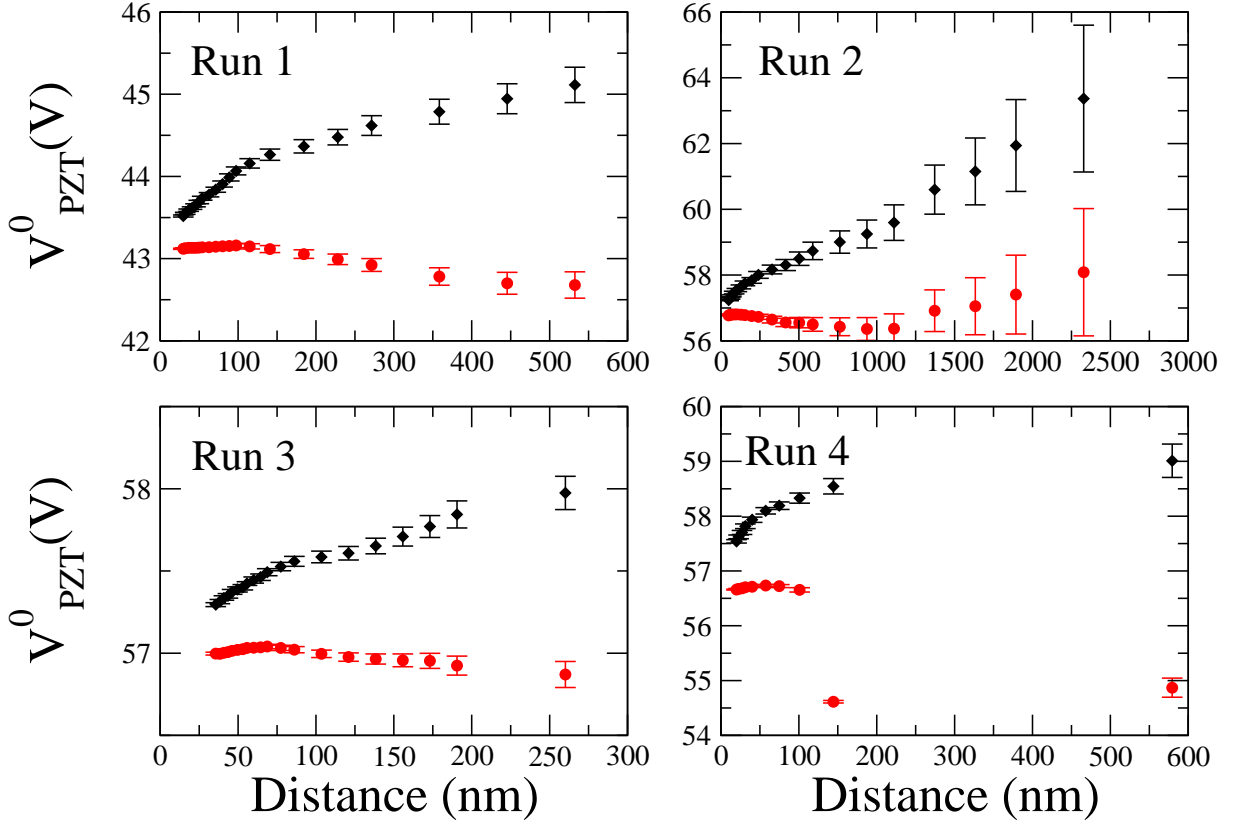}
\end{center}
\caption{Test of stability of the fit of the electrostatic
calibrations: shown is the determination from the fitting of the
offset distance $x_0=\beta V^0_\mathrm{PZT}$ versus  the
sphere- plane distance in the case of the fixed -2 exponent expected from
electrostatics (diamond points, black), and for the best exponent
left as a free parameter (circle points, red). Clearly, the trend
associated with the fixed exponent displays a wide variation in the
fitting parameter $V^0_\mathrm{PZT}$, while the overall behavior driven 
by the best exponent shows a degree of stability in all distances.}
\end{figure*}
The unexpected power law poses a significant limit on the validity of
our electrostatic calibration. Among the possible systematic 
effects causing this deviation from the expected Coulombian
behaviour, we have considered the following.

\subsection{Static deflection of cantilever}

The spring constant of our cantilever is extremely stiff (about 5400 N/m). Using
Hooke's law, a deflection experienced by the cantilever due to an
electrostatic force at 100 nm with an applied voltage of 100 mV is
less than 0.2 $\AA$. Hence, the static deflection should play little
role. 

\subsection{Thermal drift}

Even though the temperature of the cantilever is actively stabilized  by a Peltier cooler, the rest of
the system is still subject to global thermal variation.  In order to
see this, we have measured $k_{\mathrm{el}}$ with respect to time at a
nominally fixed distance. In the worst circumstance, the gap separation during
the course of measurements can drift as much as 200 nm in
{\it either} direction. Although such a drift could in principle
affect the inferred exponent, a highly unlikely non-linear
monotonic drift would be necessary to account for the consistently 
observed anomaly in each independent run.

\subsection{Nonlinearity of the PZT actuation}

The linearity of the PZT translation has been tested
under a number of different circumstances. 
Notice that the translation intervals between the data 
points in each of the runs shown in are completely random. 
Yet, all of the runs obey a specific power law in all distances. 
The PZT was also independently calibrated by means of the fiber 
optic interferometer with a consistent, linear actuation 
coefficient factor $\beta=87\pm2$ nm/V. Also, the presence of 
significant nonlinearities should manifest also as deviations from 
the PFA fitting of the sphere-plane capacitance versus the PZT
voltage. As visible in Fig. 6, this class of measurements 
strongly constraint the PZT nonlinearity. 

\subsection{Nonlinear oscillation of cantilever}

The cantilever is driven at resonance in a phase-locked loop, a routine technique
adopted by many groups \cite{Decca,Chevrierpaper,Albrecht,Giessbl}.
Higher order terms in the force expansion should produce higher
harmonics of the drive frequency. Then the assumption that the frequency
shift is simply proportional to the gradient of the external force 
could lead to erroneous assessments of  $k_{\mathrm{el}}$,
eventually affecting the exponent. We have not observed higher harmonics
in the frequency spectrum of the resonator.

\subsection{Deviation from geometrical ideality}

The deviation from geometrical ideality and its influence on the local 
capacitances \cite{Boyer} could in principle play a role, both in 
having a different shape or a variable radius of curvature, and in 
the deviations from the PFA formula. In our case, even at 
the smallest explored distances, the PFA correction due to the roughness 
is estimated to be of $0.4\%$ in the electrostatic calibration, well 
below our sensitivity.  A more careful analysis beyond the PFA is possible taking into 
account the actual surface profiles. It is however worth to point out
that at least the simple hypothesis of a non uniform radius of
curvature of the spherical surface cannot take into account the 
anomalous scaling. In PFA, the force signal (for both electrostatic 
and Casimir) is directly proportional to the radius of curvature and 
the total force will be therefore obtained as an integral along the 
radii of curvature with no deformations to the inverse square law for 
the distance dependence. In \cite{DeccaComment} an interesting
geometry has been discussed which could take into account the
anomalous exponent we have observed, however the stringent agreement 
between the capacitance measurements and the PFA expression for 
a pure sphere-plane geometry rules out this hypothesis.

\subsection{Surface roughness}

Roughness corrections become important at small separations between the surfaces,
both for electrostatic calibrations and for analyzing their residuals
in the search for Casimir forces. Usually roughness corrections in the 
electrostatic calibration are disregarded as such calibrations are 
performed at large distances. Here we consider the simplest
analysis based on the proximity force approximation. 
Let us assume that both surfaces have stochastic roughness with rms
amplitudes $\langle h_s^2 \rangle$ and $\langle h_p^2 \rangle$ for the
sphere and plane respectively. Apart from the usual PFA condition to
treat the curvature effects due to the spherical lens ($d \ll R$), we
further assume that the sphere-plane distance $d$ is 
much smaller than the lateral roughness correlation length $\xi$ on
each surface (this is the condition for the applicability of PFA to 
roughness considerations, only valid for very smooth surfaces). 
Further assuming that the rms roughness amplitudes are the smallest 
lengthscales in the problem, the PFA second-order perturbation
correction to the sphere-plane electrostatic force is

\begin{equation}
F_{\rm el}(x)= F_0(x)
\left( 1 + \frac{\langle h_s^2\rangle}{x^2} + \frac{\langle h_p^2 \rangle}{x^2} \right),
\label{rough}
\end{equation}

\noindent
where $F_0$ is the same force evaluated for smooth surfaces. 
Figure 11 shows typical images of the cantilever surface taken 
using AFM microscopy.
We measured the surface profile for our surfaces and estimate 
the rms roughness of $\langle h_s^2 \rangle$= 4 $\mathrm{nm}^2$ 
and $\langle h_p^2 \rangle$= 2.4 $ \mathrm{nm}^2$. 
The resulting deviation from ideality is below the estimated 
sensitivity of the apparatus, and an attempt to fit the electrostatic 
calibrations including the roughness at the PFA level as in Eq. 
\ref{rough} did result in significantly larger reduced $\chi^2$. 
Roughness corrections to the electrostatic calibration potentially 
represent a systematic effect that could influence the residuals 
in the search of the Casimir contribution.

\subsection{Patch potentials}
Patch effects are expected to induce deviations from the Coulombian 
scaling with distance \cite{Speake1}, and the fact that we have found 
in some calibrations a distance dependent minimizing potential could 
validate this hypothesis. However, this does not explain yet while 
the anomalous exponent is systematically observed in all runs, while 
only in some of them a manifest distance-dependent minimizing potential occurs.

\begin{figure}[t]
\vspace{1.0cm}
\begin{center}
\includegraphics[width=1.00\columnwidth]{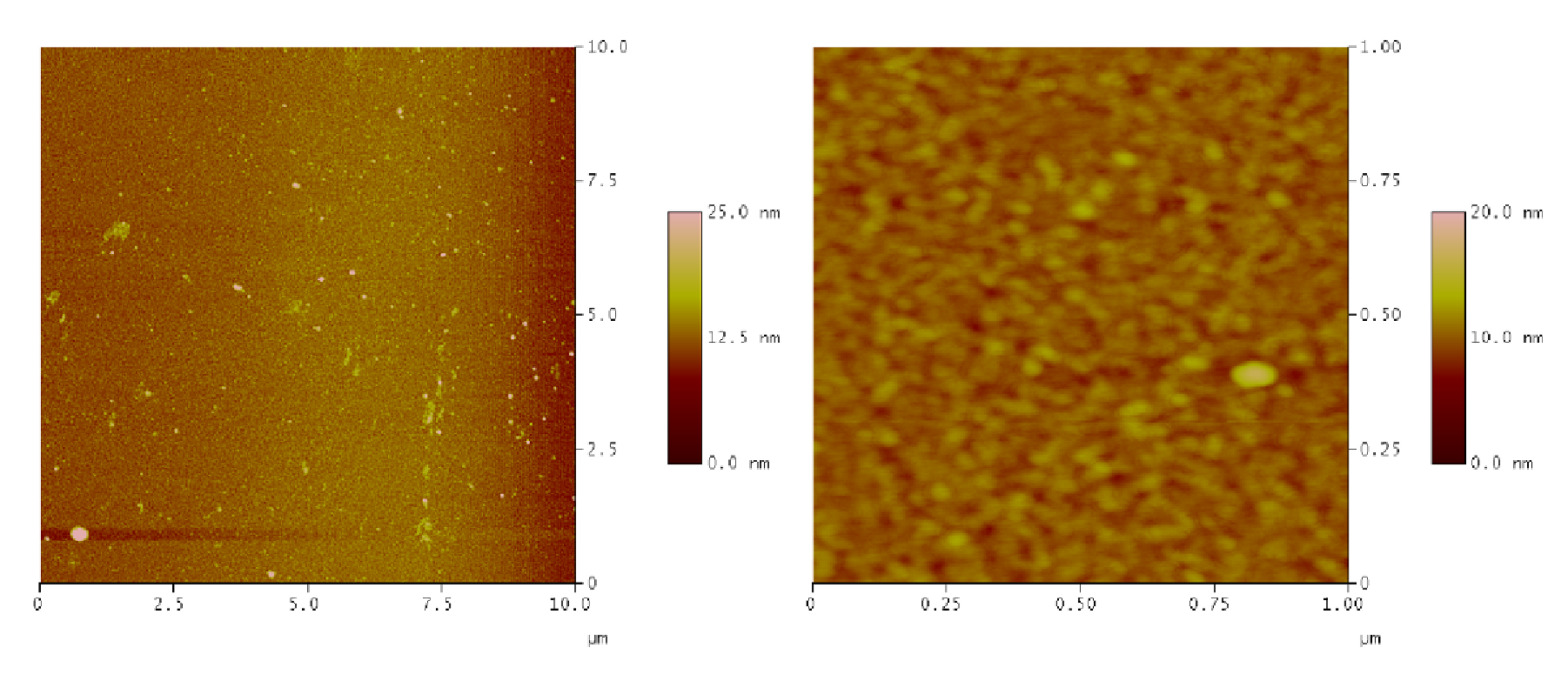}
\end{center}
\caption{Atomic force microscopy images of the cantilever surfaces, of
  size 10 $\times$ 10 $\mu$m (left) and 1 $\times$ 1 $\mu$m (right).
The overall rms roughness for the larger surface is 1.54 nm (with a
peak in the profile due to dust of 4.78 nm) and for the smaller
  surface is 0.79 nm (with a peak in the profile of 2.2 nm).}
\end{figure}

\section{Analysis of the residuals of the electrostatic calibrations
and tests of the Casimir force}

In the previous section we have found that the electrostatic calibrations performed 
in the entire measurement range show the presence of a distance-dependent 
minimizing potential (and therefore a distance-dependent contact potential) 
and an apparent deviation from the expectation for the sphere-plane capacitance
dependence on distance. While the first systematic effect can be taken
into account in the data analysis, the anomalous distance scaling of 
the electric force and the lack of its complete characterization
prevents in principle to master the calibrations.
This prevents the reliable determination of the relevant parameters 
of the system, for instance its effective mass, and therefore the
derivation of the residuals containing the Casimir force signal. 
In the absence of a complete analysis, and waiting for further
experimental and theoretical input on the anomalous exponent observed 
in our configuration, in this section we analyze the cantilever 
square frequency $\nu_0^2(x)$ obtained by subtracting to the electrostatic 
calibration data the Coulombian electrostatic contribution based on the fitting 
procedure, as first performed in \cite{Iannuzzi}. This off-line procedure is 
necessary as we have evidenced a dependence on distance of $V_0$ 
in various runs, and therefore we cannot assume {\it a priori} a 
unique counterbias voltage for an on-line subtraction of the residual 
electrostatic force as typically performed in various experiments. 
A distance-dependent residual frequency shift implies the presence of
a non-Coulombian force or drifts in the intrinsic resonator frequency
or a combination of both.  
It is evident in Eq. 8 that the electrostatic residual, 
$\Delta\nu_e^2(x,V_0)$, is zero only if the contact potential is
constant.  This term must be evaluated by first inferring
the contact potential $V_c$ through numerical integration of the
minimum condition $V_0$.

\begin{figure}[t]
\vspace{1.5cm}
\begin{center}
\includegraphics[width=0.75\columnwidth]{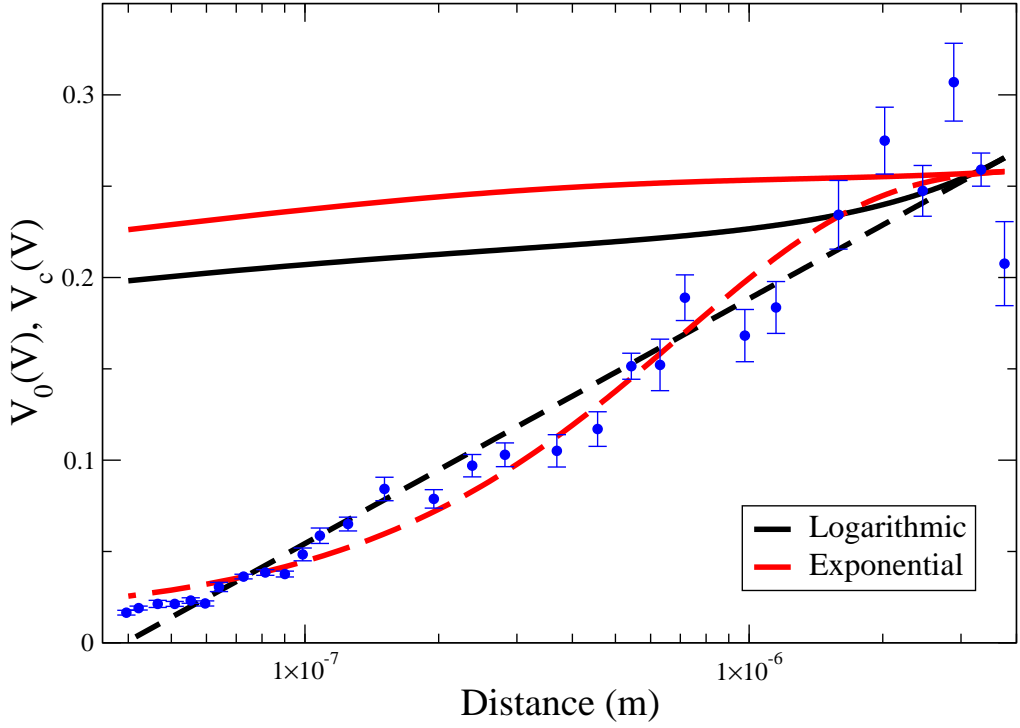}
\end{center}
\caption{Data for the minimizing potential $V_0$ versus sphere-plane
  distance for run 1. The dashed curves represent the best fits of $V_0$
  assuming an exponential dependence (red) and a logarithmic
  dependence (black). The best fit parameters are 
$V_0=(0.011 \pm 0.007)$V, $\Delta V=(0.25 \pm 0.01)$V, and $\lambda=(703 \pm 93)$ nm 
in the case of the exponential function, and $V_\mathrm{log}=(0.07 \pm 1140)$ V,
$\Delta V_\mathrm{log}=(0.058 \pm 0.003)$ V, and $\Lambda=(140.4 \pm 2.8)$ nm
in the case of the logarithmic function. The continuous curves represent the
corresponding plots of the contact potential $V_c$ after solving the
differential equation Eq. \ref{diffeq} as described in the text.}
\label{onofriobrasilia12}
\end{figure}

Based on Eq. \ref{V_0diffequ} we have a second order differential
equation for $V_c(x)$: 

\begin{equation}
\frac{C}{C''} \frac{d^2 V_c(x)}{dx^2}+2 \frac{C'}{C''} \frac{dV_c}{dx}+V_c(x)=V_0(x),
\end{equation}

\noindent
where $V_0(x)$ is determined experimentally. For the sphere-plane 
case in the PFA this results in the following differential equation:

\begin{equation}
x^2 \ln(R/x) \frac{d^2 V_c(x)}{dx^2}- 2 x \frac{dV_c(x)}{dx}+V_c(x)=V_0(x).  
\label{diffeq}
\end{equation}

\noindent
to be solved numerically for instance using a Runge-Kutta integration
method, once the experimental data on $V_0(x)$ are interpolated with 
an analytic function. In the absence of a model for the dependence on 
distance of $V_0$, different equivalent empirical functions can be
used to fit the actual data. This obviously is a strong limitation 
to the predictive power on the residuals, but at a phenomenological 
level has to be considered as a conservative and realistic procedure 
to infer some physics beyond the Coulombian contribution. 
We use, as first guess function, an exponential one such as 
$V_0(x)=V_0+\Delta V[1-\exp(-x/\lambda)]$ with $V_0$ representing the
potential when the two surfaces are in contact, $V_0+\Delta V$ is the 
asymptotic value of the minimizing potential at large distances, and 
$\lambda$ the characteristic lengthscale on which the minimizing potential varies.
Alternatively, we may use the function $V_0(x)=V_\mathrm{log} +
\Delta V_\mathrm{log} \ln(x/\Lambda)$, characterized by similar 
parameters $V_\mathrm{log}$, $\Delta V_\mathrm{log}$ and $\Lambda$,
although in this case a defined asymptotic value is not available.
In Fig. \ref{onofriobrasilia12} we show the sensitivity of $V_c$ 
obtained through integration of Eq. \ref{diffeq} to the choice of 
the function interpolating the $V_0$ distance dependence. 

Another issue requiring careful attention regards the boundary
conditions chosen to solve the second order differential equation for
$V_c(x)$. At large distances $V_c(x)$ is expected to be constant and
thus $V_c(x)=V_0(x)$. Also, the first derivative of $V_c(x)$ should be
zero in order to satisfy the differential equation. Therefore we
choose the boundary condition $V_c(x_n)=V_0(x_n); V_c'(x_n)=0$, where 
$x_n$ is the largest distance measured in the data set. 
In the case of the exponential fitting of $V_0(x)$, the curve 
becomes quickly flat as $x$ increases and therefore the use of $x_n$ 
at very large distances is a good approximation. For the logarithmic fitting 
this is more problematic, making the solution for $V_c(x)$ quite  
sensitive to the choice of $x_n$. 

Based on this input, one can evaluate now the electrostatic
contribution independent on the external bias potential, 
$\Delta \nu_e^2(x,V_0)$, and further subtract it from electrostatic
contribution dependent on the external bias evaluate at the minimizing
potential, $\nu_m^2(x,V=V_0)$ ({\it i.e.}, the maximum of each
parabola in Fig. 3). The result is shown in Fig. 13, where the dashed and
dotted-dashed curves are relative to the choice of the $V_0$
dependence on distance, as given in Fig. 12 (either exponential
or logarithmic). We see that the plotted quantities 
$\nu_m^2(x,V=V_0) - \Delta \nu_e^2(x,V_0)$ indicate that there are 
further distance-dependent residuals $\nu_p^2 + \Delta \nu^2(x)$ of 
non-electrostatic nature. In this analysis it is critical to have the 
most accurate control on the dependence of $V_0$ on distance.
In case of runs 2, 3, and 4 such a control is quite limited, and 
neither an exponential function nor a logarithmic function 
adequately describe its spatial dependence. 
In the remainder of this section we will concentrate on run 1.

\begin{figure}[t]
\vspace{1.0cm}
\begin{center}
\includegraphics[width=0.75\columnwidth]{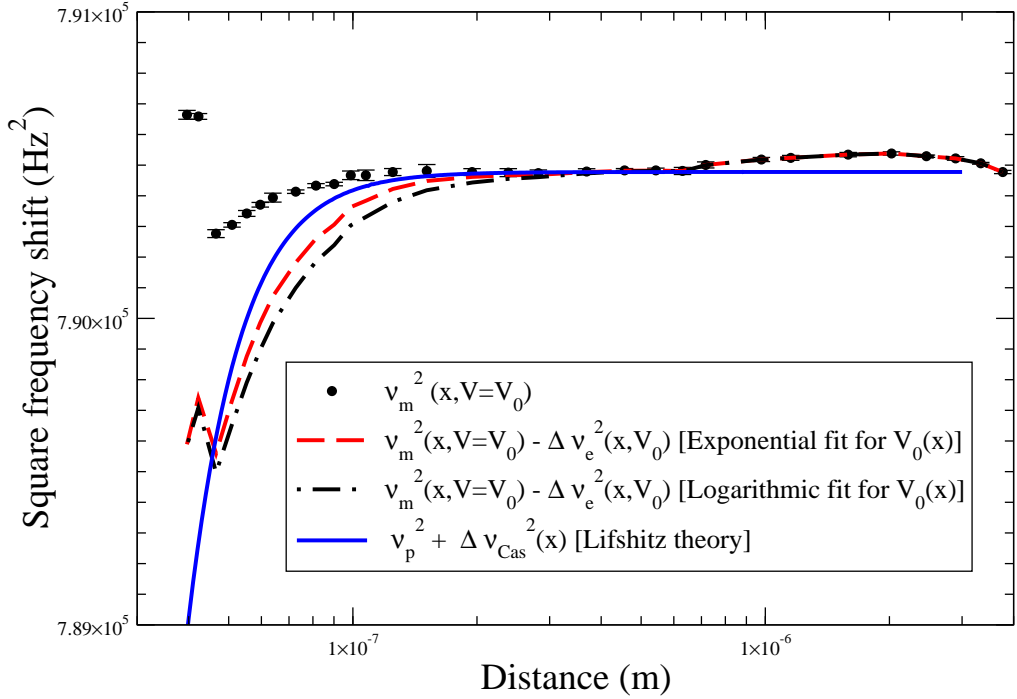}
\end{center}
\caption{
Plot of the electrostatic residuals versus distance. The dots correspond
to the maximum value of each parabola in Fig. 3 as a function of distance,
$\nu_m^2(x,V=V_0)$. The dashed (exponential fit for $V_0(x)$) and 
dotted-dashed (logarithmic fit for $V_0(x)$) lines result from
substracting from $\nu_m^2(x,V=V_0)$ the extra term $\Delta
\nu_e^2(x,V_0)$ due to the distance-dependence of
the contact potential $V_c(x)$.
The dotted line is the Lifshitz prediction for $\nu_p^2 + \Delta \nu_{\rm
Cas}^2(x)$. Parameters are an effective
mass  $m_{\rm eff}=$0.46 g  and a cantilever free oscillation frequency
$\nu_p=889.09$Hz. Optical data of gold are from Palik \cite{Palik},
and Drude parameters are $\omega_p=7.5$eV and $\gamma=0.061$eV. 
Temperature is set to $T=300$K.}
\label{onofriobrasilia13}
\end{figure}

Given the serious limitations of our electrostatic calibration, it 
is difficult to perform a rigorous analysis of these extra residuals, 
as they could be Casimir-Lifshitz forces, patch potential
forces, etc.  In the following we will {\it assume} that these 
extra residuals are only due to the Casimir-Lifshitz force, and 
evaluate the corresponding prediction for $\nu_p^2 + \Delta \nu_{\rm
Cas}^2(x)$.  As a first attempt, one can fit either the exponential 
(dashed line) or the logarithmic (dotted-dashed line) residuals in
Fig. \ref{onofriobrasilia13} with the plain Casimir formula for zero temperature and
perfect metals. In the proximity force approximation, the sphere-plane 
Casimir force is $F_{\rm Cas}(x)= 2 \pi R E_{\rm PP}(x)$, where $E_{\rm PP}(x)$ is the
Casimir energy per unit area for the plane-plane configuration, 
$E_{\rm PP}(x)= -\hbar c \pi^2 / 720 x^3$. Therefore, the frequency 
shift due to the Casimir force in the sphere-plane case is 
$\Delta \nu_{\rm Cas}^2(x) = -K_{\rm Cas} / x^4$, with 
$K_{\rm Cas} = \pi \hbar c R / 480 m_{\rm eff}$.
Fitting the above curves with  $\nu_p^2 - K_{\rm Cas}/ x^4$, we obtain
$\nu_p^2=(790440 \pm 19.3) {\rm Hz}^2$, 
$K_{\rm Cas}=(5.5164 \pm 0.3390) \times 10^{-27} {\rm Hz}^2/{\rm m}^4$
for the dashed line (exponential formula for $V_0(x)$), and 
$\nu_p^2=(790450 \pm 18.3) {\rm Hz}^2$, 
$K_{\rm Cas}=(6.5795 \pm 0.3220) \times 10^{-27} {\rm Hz}^2/{\rm m}^4$ 
for the dashed-dotted line (logarithmic formula for $V_0(x)$). 
Using the values for the radius of curvature of the sphere of
$R=30.9$mm, and an effective mass of $m_{\rm eff}=0.46$g, the 
obtained coefficients $K_{\rm Cas}$ are lower than the ideal Casimir prediction 
($K_{\rm Cas, th} = 1.3 \times 10^{-26} {\rm Hz}^2/{\rm m}^4$) by
about a factor 2, in line with the expectation that conductivity 
corrections could play a role. 

In order to analyze possible conductivity (and temperature)
corrections to the Casimir force we now use the
Lifshitz formalism \cite{Lifshitz1956}, written in terms of the
frequency-dependent reflection coefficients of the two gold surfaces.
The plane-plane free energy is
\begin{equation}
E_{\rm PP}(x) = \frac{k_B T}{2 \pi x^2} \sum_p \sum_{m=0}^{\infty '} \int_{m \gamma}^{\infty}
dy \; y^2 \; \log(1- r_p^2 \; e^{-2 y} ).
\end{equation}
Here $p$ denotes the two possible polarizations (TE and TM), $\gamma=2 \pi k_B T x / \hbar c$, and the prime
on the summation sign indicates that the $m=0$ term is counted with half weight. The reflection amplitudes
are given by the usual Fresnel formulas,
\begin{eqnarray}
r_{\rm TM}(y,\xi_m) = \frac{s_m - \epsilon(i \xi_m) p_m}{s_m + \epsilon(i \xi_m) p_m} &;&
r_{\rm TE}(y,\xi_m) =  - \frac{s_m - p_m}{s_m + p_m} ,
\end{eqnarray}
where $\xi_m=2 \pi k_B T m /\hbar$ are the Matsubara frequencies, $p_m=y/m \gamma$, and
$s_m=\sqrt{\epsilon(i \xi_m) -1 + p_m^2}$. The dielectric permittivity at imaginary frequencies is
evaluated using Kramers-Kronig relations with gold optical data from Palik \cite{Palik}, extrapolated
to low frequencies using a Drude model with parameters: $\omega_p=7.5$eV for the plasma frequency,
and $\gamma_p=0.061$eV for the plasma relaxation parameter. 
The resulting theoretical prediction (using once again PFA for the sphere-plane case)
for $\nu_p^2 + \Delta \nu_{\rm Cas}^2(x)$ at $T=300$K is plotted as
the dotted line in Fig. 13, using as the free cantilever square frequency 
$\nu_p^2 = 7.9047767 \times 10^5$ Hz${}^2$. It is apparent from the figure that the
Casimir curve $\nu_p^2 + \Delta \nu_{\rm Cas}^2(x)$ approximately follows the behavior of the curves
$\nu_m^2(x,V=V_0) - \Delta \nu_e^2(x,V_0)$ (that is, the electrostatic
residuals {\it after} substraction of the distance-dependent contact 
potential term). Note that $\Delta \nu_e^2(x,V_0) >0$, that means a
positive frequency shift associated to a {\it repulsive} electrostatic
residual ($V$-independent) force. Had we not taken this term into account, the Lifshitz
theory curve would noticeably depart from the electrostatic  residuals $\nu_m^2(x, V=V_0)$.

Finally, it is important to emphasize once again that, due to the
anomalous exponentwe found in our electrostatic calibration, we cannot make
any precision measurement claims on the Casimir force after the
residual analysis. Nor does our data have enough sensitivity to 
detect corrections due to temperature effects or 
sample dependence. On the other hand, it is quite possible that part 
of the residuals are due not only to Casimir-like  forces but also to 
other external voltage-independent forces, not taken into account in 
a simple Coulomb law, such as patch potentials.  

\section{Systematic effects and relationship to previous Casimir and
  AFM experiments}

Our experiment spans between two different classes of
experiments. In previous short-distance Casimir force experiments,
the minimum explored distance is on the order of few hundred
nanometers. This is due to the fact that typically, in order to
increase the sensitivity at larger distances, the stiffness of the
resonator is designed to be low. While this yields a stronger signal
at large distances with respect to harder resonators, this also
results in an earlier snapping of the resonator to the attracting
surface. On the other hand, typical AFM experiments aimed at
mapping the profile of a surface at subnanometer resolution
(both depth and lateral) do not need to reach distances
larger than 10-20 nm. We discuss here more in general the specific 
issues we have found that could affect at least in principle 
the analysis of previous measurements, or at least the assessment 
of their accuracy.

\subsection{Distance dependence of the contact potential}

Our finding of the dependence of the contact potential $V_c$ upon
the distance calls for a more careful study of previous experiments 
in which the contact potential was assessed at relatively larger gaps 
than the one used for looking at the Casimir force. In order to seek 
for similar regularities in previous Casimir force experiments, we 
have requested electrostatic calibration data kindly provided by
various groups. The result of the analysis for the data of the 
Lucent Laboratories group \cite{Chan} working with spheres of much 
smaller radius of curvature (100 $\mu$m) is reported in the top-right 
plot in Fig. 14, showing a small but clearly visible linear 
correlation between $V_0$ and the sphere-plane distance. 
The data of the AFM group in Grenoble (with microspheres of radius 20
$\mu$m) in Fig. 14 bottom-left show also a strong dependence of 
$V_0$ at large distances, with a more complex behavior as evident 
from the change in polarity of the minimizing potential around 5
$\mu$m. In the case of the experiment performed at Indiana
University-Purdue University Indianapolis (IUPUI), with a microsphere 
of radius (148.7 $\pm$ 0.2) $\mu$m, see bottom-right plot, no 
evident dependence on the distance can be assessed, and in that 
experiment a constant counterbias equal to the average
value was assumed. A linear trend, with a systematic difference in the contact 
potential in the explored range of about 3 mV around random
fluctuations of comparable size, seems also present in Fig. 4 of 
\cite{Mohideen08} using microspheres of radius 200 $\mu$m, although
the coarse scale used for the vertical units makes harder 
a more careful and quantitative assessment. 

Although it would be necessary to collect more data, one noticeable 
emerging pattern from the comparison of the experiments in
Fig. 14 is that experiments with intermediate radii of curvature 
generally seem to provide a milder dependence of $V_0$ within a 
comparable distance range. It is plausible that spheres with very small radius 
of curvature as in the example from the Grenoble group are more
vulnerable to local defects or impurities, while large radii of 
curvature (and therefore large active interaction surfaces) like 
in our case are sensitive to large scale variations 
in the patch potentials. It will be interesting to investigate, beside 
the obvious material composition, the geometry dependence of  
minimizing and contact potentials. An interesting study of the 
minimizing potential and its dependence upon distance and drift in time 
has been recently reported for the more macroscopic setup of a 
torsional balance \cite{Pollack}. 

\begin{figure}[t]
\vspace{1.5cm}
\begin{center}
\includegraphics[width=0.80\columnwidth]{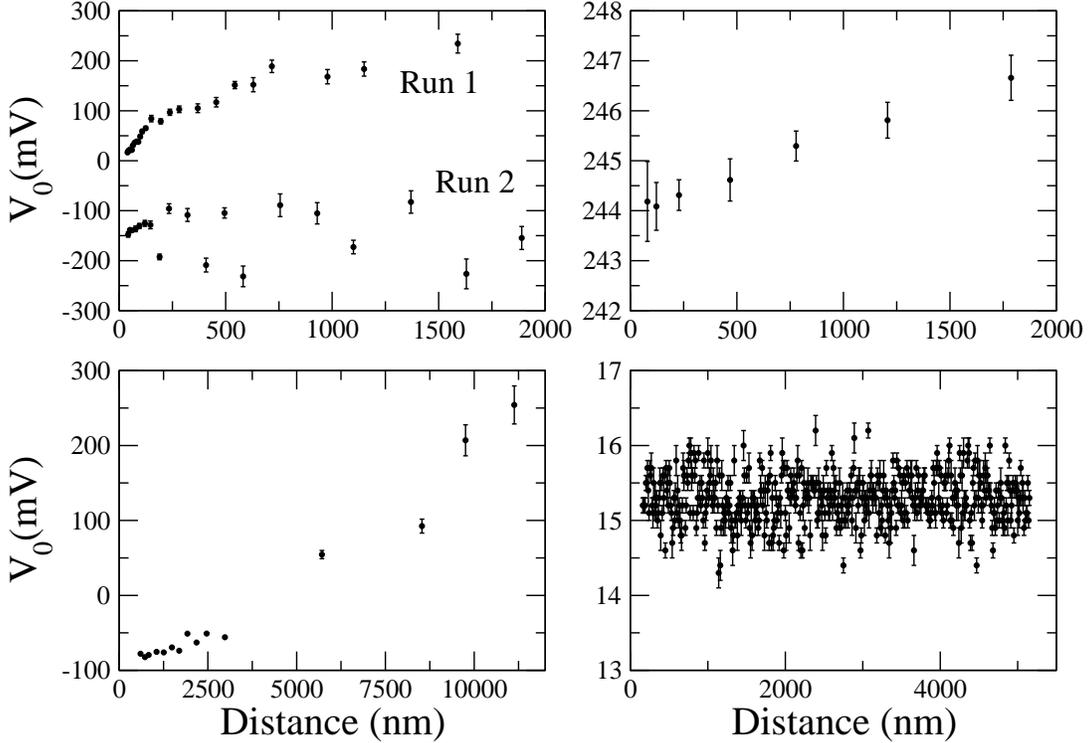}
\end{center}
\caption{Minimizing voltage versus sphere-cantilever distance resulting
for the data analysis of various experiments. 
Shown are data from run 1 and run 2 of our experiment 
(top-left, with a zero minimizing potential at the closest approach for run 1, 
and a non-zero value of $V_0 \simeq $-150 mV for the closest approach 
of run 2), data from \cite{Chan} (top-right, courtesy of F. Capasso
and H.B. Chan), results from the experiment of the Grenoble group (bottom-left, courtesy 
of J. Chevrier and G. Jourdan) , and the experiment at IUPUI (bottom-right, courtesy of R. Decca). 
The first three sets of data evidence a distance dependence of $V_0$ upon
the sphere-plane separation, which could affect significantly the
accuracy without taking into account this systematic effect, while 
this behaviour is absent in the bottom-right plot.}
\end{figure}

To assess the impact of the distance-dependent minimizing potential 
on the precision of the Casimir force measurements, following 
\cite{OnofrioCarugno}, we evaluate the equivalent voltage necessary 
to mimic the ideal (zero temperature, perfect conductors) Casimir 
force at a given distance, and in the sphere-plane case we have
\begin{equation}
V_{\mathrm{eq}}(x)= \frac{\pi}{120} \left( \frac{\hbar c}{\epsilon_0}
\right)^{1/2} \frac{1}{x} ,
\end{equation}
implying that the Casimir force can be simulated, at a distance
of 1 $\mu$m, by just having 17.5 mV of uncompensated voltage between the
two surfaces.  For real materials the Casimir force is weaker, and can 
therefore be mimicked by an even smaller uncompensated voltage.
If the minimizing potential $V_0$ is independent of the distance over
the entire range of explored distances, it is legitimate to use 
an external counterbias to cancel its effect.  
Otherwise, a fixed counterbias will only cancel out the minimizing 
potential at a given distance and, depending on the slope of $V_0$ 
versus distance at smaller gaps, most if not all of the measured 
force could be due to the uncompensated potential. Some of
the determinations of the contact potential in previous experiments
have been performed at relatively large gaps with respect to those 
at which the Casimir forces observations are reported. In some 
cases, in particular for low stiffness microcantilevers, this is 
necessary because electrostatic forces at small distances will 
cause snapping or instabilities, unless very small bias voltages, 
limited by the voltage supply specifics, are applied.
This suggests that a reanalysis of the data collected on Casimir forces 
so far will be beneficial in the light of these findings, also
including a more careful scrutiny in the next round of measurements.
Notice also that to have for instance a 0.1 $\%$ precision in the 
determination of the Casimir force the electrostatic background
should be controlled with an accuracy better than 550 $\mu$V at
1 $\mu$m, which it is also equivalent to perform precision electrometry.
In other words, the precision of 0.1 $\%$ is equivalent to be able to 
discern a minimum voltage of 550 $\mu$V applied to the gap, a very
difficult task considering the presence of comparable voltage drifts 
and patch potentials, although other groups have instead concluded 
that the estimated patch potential contributions are under control 
in their experiments \cite{Decca1,Mohideenthvsexp}. 

\subsection{Thermal expansion}

Even though the temperature of the cantilever clamping system is 
actively stabilized by a Peltier cooler, the rest of the system
is still subject to global thermal variation as large as $1$ K
during the measurement runs.
In order to understand the stability issues of our apparatus, we 
have studied the cantilever frequency for a time interval much 
longer (12 hours) than the typical duration of a run (of 3-4 hours) 
and we have show the temporal evolution of the curvature coefficient 
$K_{\mathrm{el}}$ at a given nominal distance in Fig. 8.

\begin{figure}[t]
\vspace{1.0cm}
\begin{center}
\includegraphics[width=0.70\columnwidth]{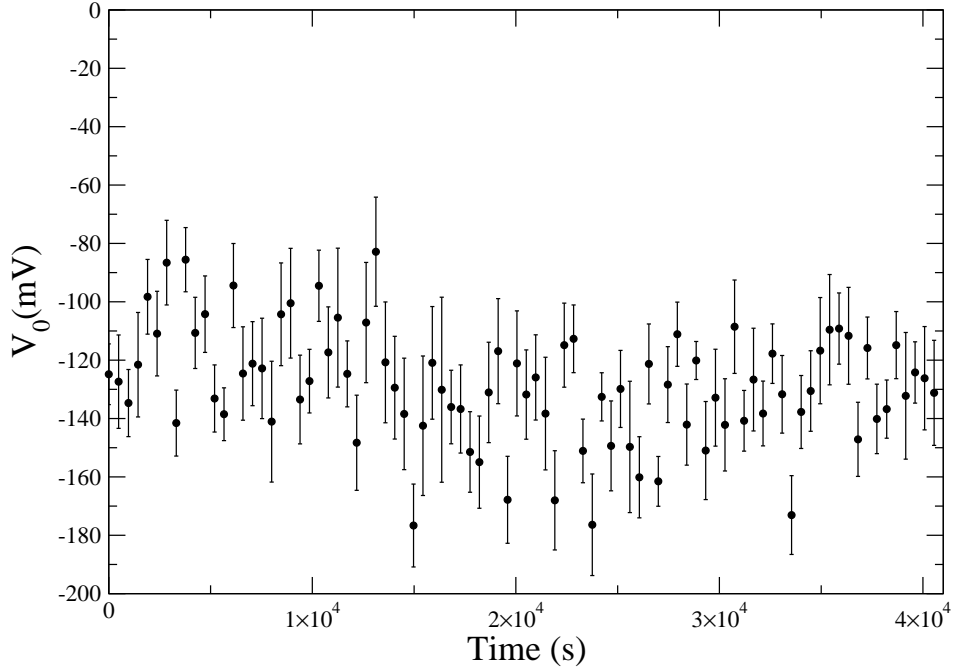}
\end{center}
\caption{Time behaviour of $V_0$ in the same long run of Fig. 8. The 
apparent lack of correlation between this drift and the one presented
in Fig. 8, which is mainly due to the drift in the sphere-plane gap
distance, seems to imply that there are genuine time-dependent drifts 
of the minimizing potential even if the sphere-plane distance is kept 
constant, consistently with the finding in \cite{Pollack}.}
\end{figure}

To understand the impact of thermal expansions let us consider 
an aluminum slab of $L=$1 cm with its coefficient of linear expansion
$\alpha=24\times10^{-6}/\mathrm{K}$, situated underneath the
cantilever. For a typical temperature change over a period of
12 hours of $\Delta T=1$K, the incremental expansion due to
temperature is then $\Delta L=\alpha L\Delta T=$240 nm, which is
almost identical as the total distance variation shown in Fig. 8. 
This translates into a range of 5-10 nm of average variation in 
the shot-to-shot electrostatic measurement, \textit{i.e.} the 
change in the absolute gap separation during the acquisition of 
an individual data point. To evaluate the robustness of the fits,
let us assume that the last position registered by the PZT right 
before hard contact is misplaced by 8 nm either closer or away from the
surface, for instance due to thermal expansion. One underlying assumption
held throughout our analysis has been the distance measurement
is solely controlled by the action of the piezoelectric transducer,
but it is evident at this point that the actual distances could
be also affected by thermal expansion. In order to see how this
small variation due to temperature change influences the final
results of the fitting procedure, we have adjusted the last
data point of the closest approach by 8 nm forward or
backward in an attempt to mimic the effect of a thermal expansion
hypothetically affecting only the last collected point.

\begin{table}[b]
\centering
\begin{tabular}{cc|cccc}
&  & Run 1 & Run 2 & Run 3 & Run 4\\
\hline
\hline
F &  $\alpha\mathrm{(Hz^2V^{-2}m^{e}})$ &$3194\pm97$ & $3200\pm158$ & $4265\pm155$ & $1749\pm 75$\\
& $x_0$ (nm) & $35.7\pm0.1$ & $53.9\pm0.3$ & $43.5\pm0.2$ & $28.7\pm0.2$\\
& $e$ & $-1.74\pm0.01$ & $-1.78\pm0.02$ & $-1.84\pm0.01$ & $-1.60\pm0.01$\\
&$\chi^2/DOF$&1.6 &0.5 & 2.2 & 7.0\\
\hline
&  $\alpha\mathrm{(Hz^2V^{-2}m^e})$ & $2805\pm92$ & $3021\pm153$ & $3732\pm144$ & $1415\pm69$\\
& $x_0$ (nm) & $29.6\pm0.1$ & $49.6\pm 0.2$ & $35.7\pm0.2$ & $20.0\pm0.2$\\
& $e$ & $-1.70\pm0.01$ & $-1.77\pm0.02$ & $-1.80\pm 0.01$ & $-1.54\pm0.02$\\
&$\chi^2/DOF$&1.0 &0.8 & 1.2 & 7.0\\
\hline
B &  $\alpha\mathrm{(Hz^2V^{-2}m^e})$& $2788\pm97$ & $3006\pm162$ & $3608\pm149$ & $1451\pm78$\\
& $x_0$(nm) & $28.7\pm0.2$ & $48.7\pm0.3$ & $33.9\pm0.2$ & $20.0\pm0.2$\\
& $e$ & $-1.70\pm0.01$ & $-1.77\pm0.02$ & $-1.79\pm0.01$ & $-1.54\pm0.02$\\
&$\chi^2/DOF$&2.8 &1.6 & 1.7 & 11.5\\
\hline
\hline
\end{tabular}
\label{fitadjust}
\caption{Stability test of the fitting parameters for the electrostatic
calibrations. The middle rows present the fitting parameters
$\alpha$ and $x_0$ for the electrostatic calibrations
with a power law exponent $e$ taken to be a free parameter, such that
$K_{\mathrm el} \propto d^e$. F and B in the first and third rows
represent hypothetical  situations in which the last data point right
before hard contact were moved by 8 nm forward (F) and backward (B)
from the original position, respectively.}
\end{table}

The relocation of the single data point of the final distance has 
modified the entire set of fitting parameters as shown in Table 2. 
For the case of the calibration factor $\alpha$, variations due to 
the modification of the single last data point were $14\%, 6.4\%,
18\%, 21\%$ for Run 1-4, respectively. Even larger changes in the 
values of the distance offset $x_0$ were found, up to 40 $\%$.

The exponent $e$ of the electrostatic power law, however, remains
stable over the forward and backward displacements of the single data
point, although its overall variation is evidently greater than the
fitting uncertainties at a given position, demonstrating how
vulnerable fitting parameters are to small changes in the absolute gap
distance. It should be emphasized that only a single data point is
readjusted in this analysis. Because the thermal expansion is present
throughout the whole measurement at all distances, what has been shown
here only represents a subset of many other possibilities.

Another source of drift has been identified in the minimizing
potential, as evidenced in Fig. 15. This is consistent with recent 
reports of dedicated experiments studying the long term drift 
of the surface potential using torsional balances \cite{Pollack}.
The presence of a minimizing potential slowly drifting in time 
implies severe limitations on the possibility to integrate for 
long time intervals precision force measurements. 

\subsection{Relationship between Casimir and AFM measurements}

For the sphere-plane geometry, Casimir force experiments have used
sphere radii much larger than typical in AFM apparatus ($R \simeq
100\mu$m versus 10-100 nm respectively) due to different motivations 
(mesoscopic effective surface versus higher image resolution, respectively). 
However, the physics ruling the instrument should be the same and one
expects a smooth transition between the two regimes, and a close
comparison between the accuracies reachable in these two regimes. 
Many of the systematic effects that can potentially mimic the Casimir 
force have been extensively studied in the AFM experiments 
\cite{Binnig,Garcia,Giessbl}, in particular for the investigation 
of adhesion and friction surface forces 
\cite{Trappedcharge,Electrification,Blackman,Anisotropy}, 
many years before Lamoreaux's landmark experiment \cite{Lamoreaux}.
D\"{u}rig {\it et al.} have studied metallic adhesion and short range
forces using Scanning Tunneling Microscopy (STM)
\cite{Durig1990,Durig1994}, while a number of papers are devoted to
the subject of radiative transfer \cite{Hargreaves}, noncontact
friction \cite{Volokitin}, and dissipative interaction between a
resonant cantilever and various sample surfaces in the dynamic AFM
mode \cite{Stipe,Dorofeyev,Fuchs,Loppacher,Dissipation,Ueba}, still an
active topic of research. It would be productive to bridge the 
short-range AFM measurements and Casimir force measurements that, 
after the pioneering measurements of the macroscopic van der Waals 
force \cite{Tabor2}, seem to have developed with somewhat divergent 
methodologies.

The Casimir force may be considered as the van der Waals force in the
long-range, retarded regime and therefore the study of its nature may
be seen as an extension of the general studies of the previous van der
Waals force measurements with AFM techniques. From this perspective,
the measurement of forces between macroscopic surfaces at small
distances becomes a rich field of study, including also the possible
role played by the meniscus force, double-layer force, capillary
force, hydration force, hydrophobic force, and steric, depletion, and
bridge forces, all ultimately of electromagnetic origin, yet different
in distance scaling and strength \cite{Cappella}. Although not all of
these forces may be relevant in a given experimental situation, one
must carefully check the presence of each of these forces and must be
able to distinguish them, on the top of residual electrostatic forces
and Casimir forces in themselves. It is plausible that, especially 
at the level of accuracy reachable in current and future high precision 
measurements of the Casimir force at small distances, some of these 
forces should appear already at the level of residuals from the best fits.

A similar argument may be also applied to the electrostatic
calibrations.  The AFM community has already discussed issues such as
friction, electrification, patch charges and potentials, while these
have received little attention so far in the Casimir framework (apart
from the estimates for the effect of the patch potentials discussed in
\cite{Decca1,Mohideenthvsexp}), in spite of relevant analysis
appearing in the recent literature \cite{Speake1,Robertson,Palacios}. 
Also, Burnham {\it et al.} \cite{Anisotropy} have evaluated the effect
of patch charges by means 
of the image method, and compared with the previously observed long-range
interaction of van der Waals force in the surface force microscopy. In
\cite{Stipe}, the force fluctuations between a cantilever tip
($R=1\mu$m) and a surface (both gold-coated) are interpreted in terms
of inhomogeneous electric fields due to the presence of atomic steps,
adsorbates, hillocks, pits, and other defects. The ubiquity of these
electric fields may significantly limit precision and accuracy of Casimir force 
measurements. It is therefore of great relevance to aim also at 
calibrations with other physical effects, such as through radiation 
pressure \cite{Petrov} and hydrodynamic forces in the case of the 
Casimir-Lifshitz force in the presence of liquid media \cite{Munday}, 
and to perform detailed comparisons among the various calibration
techniques, especially to filter out issues specific to the
electrostatic calibrations.

\section{Conclusions}

We have discussed electrostatic calibrations in a sphere-plane setup 
different from the previous ones for two features, namely a large 
radius of curvature of the sphere and a microresonator with large
stiffness. This combination allows us to extend electrostatic 
calibrations to small gap distances in the presence of large 
spherical surfaces. We have evidenced anomalies in the expected 
scaling of the electrostatic force, and a distance dependent 
potential minimizing the electrostatic contribution. In the 
absence of a complete control of the underlying electrostatic 
physics, it is difficult to calibrate the apparatus at the extent to 
meaningfully discuss the residuals signal. By using the electrostatic 
expectations, we have found evidence of the Casimir-Lifshitz force at
small separation in one out of the four runs for which optimal working
conditions were found and a complete data analysis have been performed. 
While we believe that some systematic effects are being emphasized by
using spheres with large radius of curvature and small sphere-plane 
distances, we draw two conclusions from our results which may be of
more general interest. Firstly, we have shown that the determination at all
distances of the contact potential $V_0$ is crucial, and its
uncertainty can affect the entire data analysis procedure. 
This implies that electrostatic calibrations have to be performed 
in the entire range of explored distances, rather than being limited to
larger separations for which the Casimir force is expected to be
negligible. Secondly, we have discussed the stability of the 
determination of the fit parameters, like distance offset, minimizing 
potential, and effective mass, coming from the electrostatic
calibrations. Various groups are now trying to confirm or rule out the 
anomalies we have observed in our experimental setup \cite{IannuzziNew}. 
Apart from pointing out some limitations of the sphere-plane geometry,
we believe that our discussion of systematic effects 
and data fitting robustness is beneficial for more rigorous data analysis 
in the next generation of experiments for any geometrical
configuration, allowing to explore non-Coulombian forces, in 
the spirit of an experiment able to discover unknown physics rather 
than the demonstration of an {\it a priori} known effect \cite{Onofrio}.

\vspace{0.1cm}
\noindent
$^*${Present address: Department of Physics, Yale University, 217
  Prospect Street, New Haven, CT 06520-8120, USA}

\vspace{0.4cm}
\noindent
We thank R.L. Johnson for skillful technical support, and 
Q. Wei for experimental assistance. M.B.H. acknowledges support 
from the Dartmouth Graduate Fellowship program and the NSF GAAN 
program. W.J.K. acknowledges useful discussions with S.K. Lamoreaux and 
A. Parsegian, while D.A.R.D. acknowledges interesting discussions 
with G. Carugno, H.B. Chan, R. Decca, S. de Man, S.K. Lamoreaux, U. Mohideen, 
J. Munday, and G. Ruoso. 
R.O. acknowledges partial support by the NSF through the Institute 
for Theoretical Atomic and Molecular Physics at Harvard University 
and the Smithsonian Astrophysical Observatory, and is grateful to 
F. Capasso and H.B. Chan for kind permission to use a set of their 
electrostatic calibration data from \cite{Chan} for the analysis 
reported in Fig. 14, to J. Chevrier and G. Jourdan for useful 
discussions and for providing data reported in Fig. 14 prior 
to their publication, and to R. Decca for useful discussions 
and for providing data of his experiment reported in Fig. 14.

\end{document}